\documentclass[journal]{IEEEtran}
\usepackage{cite}

\ifCLASSINFOpdf
\else
\fi
\usepackage{amsmath}
\usepackage{algorithmic}

\usepackage{array}

\ifCLASSOPTIONcompsoc
  \usepackage[caption=false,font=normalsize,labelfont=sf,textfont=sf]{subfig}
\else
  \usepackage[caption=false,font=footnotesize]{subfig}
\fi
\usepackage[ruled,vlined]{algorithm2e}
\usepackage{amssymb}
\usepackage{amsfonts}
\usepackage{amsthm}
\usepackage{mathtools}
\usepackage{booktabs}
\usepackage{color}
\usepackage{epsfig}
\usepackage{multirow}
\usepackage{cite}
\usepackage[hyphens]{url}

\newcommand{\argmin}{\operatorname*{arg\,min}}

\newcommand{\dc}{\mathsf{x}}
\newcommand{\etal}{\textit{et~al.}}

\usepackage[utf8]{inputenc}
\usepackage{siunitx}
\newcommand{\RN}[1]{%
	\textup{\uppercase\expandafter{\romannumeral#1}}%
}
\newcommand{\tablerowstretch}{1.2}

\begin{document}
%
\title{Depth Sequence Coding with Hierarchical Partitioning and Spatial-domain Quantisation}
%
%
%

\author{Shampa~Shahriyar,
        Manzur~Murshed,
        Mortuza~Ali,
        and~Manoranjan~Paul
    \thanks{S.~Shahriyar was with the Faculty of Information Technology, Monash University, Victoria, Australia (e-mail: shampa077@gmail.com).}%
    \thanks{M.~Murshed is and M.~Ali was with the Faculty of Science and Technology, Federation University Australia, Churchill Vic 3842, Australia (e-mail: manzur.murshed@federation.edu.au; mortuza94@gmail.com).}%
    \thanks{M.~Paul is with the School of Computing and Mathematics, Charles Sturt University, Bathurst NSW 2795, Australia (e-mail: mpaul@csu.edu.au).}%
	\thanks{This research is supported by the Australian Research Council Discovery Project DP130103670.}}%
\maketitle

\begin{abstract}
	Depth coding in 3D-HEVC for the multiview video plus depth (MVD) architecture (i) deforms object shapes due to block-level edge-approximation; (ii) misses an opportunity for high compressibility at near-lossless quality by failing to exploit strong homogeneity (clustering tendency) in depth syntax, motion vector components, and residuals at frame-level; and (iii) restricts interactivity and limits responsiveness of independent use of depth information for ``non-viewing'' applications due to texture-depth coding dependency. This paper presents a standalone depth sequence coder, which operates in the lossless to near-lossless quality range while compressing depth data superior to lossy 3D-HEVC. It preserves edges implicitly by limiting quantisation to the spatial-domain and exploits clustering tendency efficiently at frame-level with a novel binary tree based decomposition (BTBD) technique. For mono-view coding of standard MVD test sequences, on average, (i) lossless BTBD achieved  \SI{42.2}[\times]{} compression-ratio and \SI{-60.0}{\percent} coding gain against the pseudo-lossless 3D-HEVC, using the lowest quantisation parameter $QP = 1$, and (ii) near-lossless BTBD achieved \SI{-79.4}{\percent} and \SI{6.98}{\dB} Bjøntegaard delta bitrate (BD-BR) and distortion (BD-PSNR), respectively, against 3D-HEVC. In view-synthesis applications, decoded depth maps from BTBD rendered superior quality synthetic-views, compared to 3D-HEVC, with \SI{-18.9}{\percent} depth BD-BR and \SI{0.43}{\dB}  synthetic-texture BD-PSNR on average.
\end{abstract}


%
\IEEEpeerreviewmaketitle

\section{Introduction}
\label{sec:Ch3_intro}

\IEEEPARstart{M}{ultiview} video is an emerging trend in the development of interactive digital video systems to offer immersive viewing experiences. Acquisition and transmission of multiview video content is constraint by the available storage and bandwidth. The bitrate requirement for transmitting encoded multiview video content increases almost linearly with the number of views (aka viewpoints). To circumvent this only a limited number of views, spaced evenly in terms of the viewing-angle, are encoded and virtual-views for intermediary viewpoints are synthesised on-demand from the adjacent encoded-views using the depth image based rendering (DIBR) techniques~\cite{Paper10_Fehn_DIBR_SPIE2004}. The essential idea of DIBR is to extrapolate the texture, a collective term referring to the luminance (intensity) and chrominance (colour) values of pixels, in a virtual-view from the texture and associated depth map, representing the distance of pixels from the corresponding points on the surfaces of objects in the \num{3}-dimensional (3D) scene, in the encoded-views that are adjacent to the virtual-view.

\bstctlcite{my:BSTcontrol}

Compared to texture, depth map typically exhibits distinct structural properties with higher homogeneity and sharp edges at object boundaries. Consequently, traditional video coding techniques render ineffective to compress depth map sequences at the high ratio expected from exploiting homogeneity. To address this, 3D-HEVC~\cite{sullivan2013standardized,tech2016}, the multiview extension of the latest High Efficiency Video Coding (HEVC) standard, has adopted the \emph{multiview video plus depth} (MVD) architecture. In contrast to video (texture) coding, 3D-HEVC encodes depth map sequences with fewer and additional coding paths/modes to exploit the spatio-temporal correlations at the intra-view, inter-view, and inter-domain (texture-depth) levels.

Depth map compression techniques \cite{Paper74_Wiegandicme2013,Paper3dhevc2013,Paper80_Zamarin_icme2013} adopted in 3D-HEVC, in general, exploit smooth-regions at block-level. A coding-unit block, which partially covers two or more regions, is partitioned into homogeneous segments by approximating (i) the boundaries with wedgelets or contours and (ii) the segments with constant partition values (CPVs). However, the potential compression efficiency from exploiting higher homogeneity in depth maps is too great to extract at block-level. Not only is an opportunity to encode smooth-regions as-a-whole at frame-level missed but noises are also introduced at the sharp boundaries, the most-sensitive components of depth maps. Consequently, depth map coding in 3D-HEVC suffers from the following three shortcomings that curb the potential use of depth data in much wider applications.

\emph{Firstly}, the lossy techniques \cite{Paper74_Wiegandicme2013,Paper3dhevc2013,Paper80_Zamarin_icme2013} obscure sharp edges due to edge-approximation, spatial- to frequency-domain transformations, and aggressive quantisations~\cite{shampadicta}. As a result, noticeable shape deformations are introduced at object boundaries in the synthesised views rendered from the decoded depth maps of 3D-HEVC. Consequently, overall view-synthesis quality is compromised.

The proposed motion-compensated depth map sequence coder preserves edges \emph{inherently} by limiting quantisation to the spatial-domain with a small scalar step size on the pixel-level residual values. Moreover, the decoded depth maps are near-lossless. On standard MVD test sequences, it has retained very high quality with average peak signal-to-noise ratio (PSNR) of \SI{52.2}{\dB} for the recommended maximum quantisation step size $Q = 15$. Retaining such high quality in the decoded depth maps may be explained as follows. In audio and image coding, prediction residuals, rounded to the nearest integers, tend to follow two-sided geometric (TSG) distribution~\cite{Ali2006}, modelled as $\Pr(r = k) = ((1 - p)/(1 + p))^{|k|} p$, $k \in \mathbb{Z}$, where parameter $p$ is the proportion of zeros. For any arbitrary quantisation step size $Q = 2D + 1$, the mean squared error (MSE) is estimated as 
\begin{equation}
\text{MSE} = 2 p \sum_{k = 1}^D k^2 \sum_{i = 0}^{\infty} \Bigl( \frac{1-p}{1+p} \Bigr)^{i (2D + 1) + (-1)^i k}
\end{equation}
by considering the symmetry in squared errors about the $y$-axis. Depth residuals typically have very high proportion of zeros e.g., $p \in \{0.8,0.9\}$. For $Q = 15$, the MSE is typically expected to be in the range \numrange[range-phrase = --]{0.117}{0.282} with PSNR $20 \log_{10} (255/\sqrt{\text{MSE}})$ in the range \SIrange{52.63}{57.44}{\dB}.

\emph{Secondly}, quality of decoded depth maps by 3D-HEVC, even at low quantisation, is not acceptable for many ``non-viewing'' applications such as  auto-navigation, night vision imagery, object tracking, action recognition, and many computer vision applications. 

Efficient near-lossless (or even lossless) compression of depth maps can provide an attractive alternative to lossy coding, since little (or no) distortion is introduced in the synthesised views by the depth map coder. For natural images (texture), lossless coding typically attains compression ratio of merely \SI{2}[\times]{} to \SI{3}[\times]{} that prohibits any bitrate-sensitive applications. A much higher compression ratio is achievable for lossless coding of depth maps by exploiting the prevalent high level of spatio-temporal homogeneity. At lossless mode, the proposed coder has achieved average compression ratio of \SI{42.2}[\times]{} on multiview standard test sequences. The compression ratio at near-lossless mode, which further applies modest spatial-domain quantisation, is so high (on average \SI{608.8}[\times]{} while introducing negligible distortion at PSNR \SI{52.2}{\dB}) that the need for any lossy depth map sequence coder is no longer justified. 

Achieving such high compression ratio for depth map sequences without incurring significant distortions may be explained as follows. Motion-compensated depth-residuals are mostly low values, or simply zeros, due to the prevalent high temporal correlation in successive depth frames, except in the moving regions where noticeable variance in depth may be observed. Intra-predicted depth-residuals also exhibit dominant low values due to the high level of spatial correlation in depth values from the same object in the 3D scene. On top, the low dynamic range in typical depth signals makes sure that a large proportion of a depth residual frame has very low magnitude values after applying modest spatial-domain quantisation. Ultimately, high \emph{clustering tendency} is exhibited in depth residual frames as well as the corresponding syntax (coding tree unit divisions and coding unit modes) and motion ($x$- and $y$-components) frames where regions with highly skewed distribution of values emerge. This unique property is exploited by the proposed coder with a novel \emph{hierarchical partitioning} technique for frame-level data maps to achieve significant compression efficiency gain by encoding each partition separately using \emph{arithmetic coding}, which is particularly effective on very skewed probability distribution.

\emph{Thirdly}, the texture-depth coding dependency in 3D-HEVC restricts interactivity and limits responsiveness of independent use of depth information for non-viewing applications as a depth map cannot be decoded without decoding the corresponding texture frame. Moreover, the inter-view coding path, where depth maps in already-encoded views are used as references to compress the depth map of another view, curbs random access functionality and restricts interactivity.

The paper develops an independent (standalone) depth map sequence coder, which can operate both at lossless and near-lossless mode by appropriately setting the frame-level scalar quantisation step to zero and a small positive value, respectively. High clustering tendency in the frame-level data maps are exploited efficiently with a hierarchical partitioning technique, namely \emph{binary tree based decomposition} (BTBD). Using a code-length estimator of  arithmetic coding, BTBD splits a data map recursively at a hyperplane orthogonal to one of the axes using a greedy optimization heuristic. Effectively, it divides the data maps into relatively-homogeneous cuboids of arbitrary sizes that are encoded independently with overall compression efficiency significantly higher than that without partitioning. As the level of distortion in a decoded depth frame depends solely on the quantisation step size used for the corresponding residual frame, the rate-distortion optimisation (RDO) is simplified to consider only the code-length of different coding modes. Being independent from texture coding, the proposed depth sequence coder is highly responsive (fast random access functionality) and flexible (can be used with coders of other modalities) to offer wider applications of depth data.

Effectiveness of spatial-domain quantisation in inherently preserving edges in depth maps was first demonstrated in~\cite{shampaicme,shampadicta} where 3D-HEVC depth residuals were quantised at pixel-level and each frame is encoded in whole with lossless JPEG (JPEG-LS)~\cite{JPEGLS}, the lossless compression scheme developed by the Joint Photographic Expert Group (JPEG), to exploit long 1D-runs of very low values. The preliminary concept of BTBD was first introduced in~\cite{shampadcc,7457774} to exploit the prevailing clustering tendency in 3D-HEVC depth coding modes (2D-runs) and depth motion vector components (3D-runs). Full potential of these novel ideas, however, cannot be realised within the 3D-HEVC framework. Firstly, its edge-approximation coding modes contradict the philosophy of inherent edge preservation. Secondly, spatial-domain quantisation renders its RDO ineffective or, at best, sub-optimal. Finally, the it has no provision for lossless depth map coding.

This paper presents a comprehensive depth map sequence coder, independent from the 3D-HEVC framework, with the following key contributions: (i) comparative analysis of clustering tendency in motion-compensated texture and depth residual frames; (ii) in-depth analysis of achievable depth map quality with spatial-domain quantisation; (iii) rank-based signed residual mapping; (iv) RDO-based coding tree unit (CTU) division and coding unit (CU) mode selection; (v) formation of 2D/3D data maps to exploit clustering tendency in depth syntax, motion, and residual frames; (vi) enhanced context modelling for context-adaptive arithmetic coding (CAAC) of BTBD partitions; and (vii) comprehensive performance analysis of BTBD at lossless and near-lossless modes. Some preliminary results of this work, limited to only lossless mode with a fixed size CUs (\SI[product-units=single]{8 x 8}{pixel}) and approximated bitrate (using estimated code-length of arithmetic coding), have been published recently in~\cite{7552984}.

Rest of the paper is organised as follows. In Section~\ref{sec:ch3_relatedwork}, state-of-the-art depth map coding techniques are reviewed. Clustering tendency in depth maps is analysed in Section~\ref{sec:ch3_uniqueobservation} to provide motivation of the proposed hierarchical partitioning based coding scheme, which is elaborated in  Section~\ref{sec:Ch3_depthcoder}. Section~\ref{sec:Ch3_result} presents simulation results with analyses and  Section~\ref{sec:Ch3_conclusion} concludes the paper.

\section{Related Work}
\label{sec:ch3_relatedwork}

Both lossy and lossless depth map coding schemes have been proposed in the literature.

\subsection{Lossy coding of depth maps}

Depth map coding schemes proposed in the literature can be broadly categorized into three classes: segmentation, block partitioning, and edge-adaptive transformation (EAT) based approaches.

\subsubsection{Segmentation based approach} 

The schemes proposed in~\cite{zhu2009view,milani2010depth,Paper77_Farbian2011,Paper3_SungHo_3DRExpress2013} explored the idea of explicitly identifying object boundaries from depth maps using segmentation techniques. These schemes then aim at separately encoding the object boundaries followed by compression of the smooth regions. In this approach, the overhead of explicitly coding the contours outweighs the gain achieved by efficient compression of the smooth regions, affecting the overall coding efficiency. Another important limitations of this approach is that the explicit segmentation process is computationally expensive.

\subsubsection{Block partitioning based approach}

Alternative schemes that avoid explicit object segmentation have also been proposed in the literature~\cite{Paper37_Morvan2006,Paper38_Morvanicip_2007,Paper39_lucas_icip2012}. The essential idea is to represent a frame using a quad-tree partition where object boundaries at leaf nodes are modeled using simpler geometric primitives. Morvan \etal~ proposed platelet-based depth coding in~\cite{Paper37_Morvan2006, Paper38_Morvanicip_2007}. It assumes that a leaf node in the quad tree is piecewise planar and thus can be modelled with two regions of constant gradient separated by a straight line. Thus, the scheme uses piecewise-linear functions (platelets) to approximate the edges in the blocks. While this scheme partitions and models the original depth maps, the scheme proposed in~\cite{Paper39_lucas_icip2012} models the residual, resulting from intra-prediction, using  platelets. This scheme was later extended by Merkle \etal~in \cite{Paper74_Wiegandicme2013}. It proposed using wedglet, instead of platelets, to model the residual, which has been adopted in the latest 3D-HEVC standard~\cite{Paper3dhevc2013, sullivan2013standardized}. 

Zamarin \etal~\cite{Paper80_Zamarin_icme2013} introduced a new intra-mode, specifically targeted to depth macroblocks with arbitrarily shaped edges. The scheme also partitions the edge-macroblock into two regions, each approximated by a flat surface.  Considering the edge structure of previously encoded macroblocks as the context, the scheme proposed context based encoding of the edge information. To obviate the necessity of transmitting the edge information, Kang \etal~\cite{Paper73_kang2010} proposed partitioning a macroblock using the edge estimated from the geometric structure of the previously encoded neighboring blocks. 

These block-based depth coding techniques mostly use intra-coding that cannot take the advantage of high temporal correlations in successive depth maps and spatial correlations in inter-coded depth residuals. Besides, their handling of edges by approximating them with piecewise linear functions essentially distorts the shape of the object.

By assuming significant statistical correlation between texture and depth maps, many techniques tried to reuse block partitioning information for depth coding from coded texture images by recognizing the structural similarity between the texture image and the corresponding depth map~\cite{Paper3dhevc2013,Liu2011,Merklepcs2012}. Merkle \etal~\cite{Paper74_Wiegandicme2013} have also introduced contour based depth edge modelling using information from the corresponding texture image. The assumption that an edge in texture image is reflected on the corresponding depth map does not always hold in practice. If an object has textured patterns, abrupt discontinuity of colour values appear inside as well as at the object boundary in the texture image. However, the corresponding depth map exhibits discontinuity of depth values only at the boundary as the inner part of the object is represented by a constant depth value.

There are also several methods~\cite{sungho2006,Paper1_Grewatsch2004,hannuskela2012,liu2011spie} that take the advantage of shared motion information (generated from block based motion search~\cite{motionsearch2000}) between texture and depth data for coding efficiency. Due to lighting condition changes, motion compensation are not similar for texture and depth. While searching for a best matching block, texture motion search can result in long motion vectors and large residual values. On the other hand, corresponding depth block is more similar to co-located or neighbouring blocks in the reference frame. Thus, reuse of texture motion vectors is not guaranteed to achieve optimal rate-distortion performance~\cite{shampadicta}.

\subsubsection{EAT based approach}

Maitre and Do~\cite{shapewavelets} proposed a depth map compression scheme based on a shape-adaptive wavelet transform by generating small wavelet coefficients along depth edges. Shen \etal~\cite{Paper40_ortegaedgetransform2010} proposed a set of EATs to replace the standard discrete cosine transform (DCT). This block based scheme essentially (i) detects the edges in a block; (ii) constructs a graph based on the edge map of the block, and (iii) computes an EAT for the graph that minimizes a number of non-zero coefficients that must be encoded for the block. Motivation for this approach stems from the fact that for piecewise smooth signals e.g., depth maps, EAT would yield sparser representation. 

Compressed sensing (CS) is an emerging area in signal processing, which suggests that any sparse signal can be reconstructed from the sub-samples using efficient recovery algorithms. Recently, depth map coding schemes~\cite{duan2011improved, do2010compressive,lee2012adaptive} have been proposed under the CS framework. By assuming that image blocks are sparser in the pixel domain than in the residual domain, Do \etal~\cite{do2010compressive} and Duan \etal~\cite{duan2011improved} proposed CS method based on down-sampling of the 2D-DCT coefficients. However, the choice of DCT as sparsifying basis is inefficient for blocks containing arbitrarily shaped edges. Therefore, a CS based scheme using EAT as sparsifying basis has been proposed in~\cite{lee2012adaptive}.

One of the shortcomings of the EAT based approaches is that they require explicit edge detection and lossless transmission of the edge map to the decoder so that the decoder can also construct the same EAT. Besides, the process of constructing the EAT is computationally expensive.

\subsection{Lossless coding of depth maps} 

Very few works on efficient lossless compression of depth maps have so far been proposed in the literature. Kim \etal~proposed a bit-plane based scheme in~\cite{Paper55_park2010}. The scheme proposed to decompose the original frame into several bit planes by first transforming the depth values using  gray code. Then, the decomposed bit planes were encoded independently and in-order, starting from the MSB bit plane to the LSB. The encoding process is thus repeated eight times. Zamarin and Forchhammer~\cite{Paper81_Zamarin_icme2012} later modified this scheme by changing the prediction template (inter and intra) and extending the prediction into view level.

Since CABAC (context-adaptive binary arithmetic coding)~\cite{CABAC} was originally designed for lossy texture coding, it was unable to provide the optimum coding performance for lossless depth map coding. In~\cite{improvedcabac}, Heo and Ho proposed an improved version of the arithmetic encoder of H.264/AVC using significance map targeting depth map encoding. The technique, however, used only fixed \SI[product-units=single]{4 x 4}{pixel} blocks.

\section{Clustering Tendency in Depth Maps}
\label{sec:ch3_uniqueobservation}

In this section, a comparative analysis of clustering tendency in motion-compensated texture and depth residual frames is briefly presented. Let $I_{\text{texture}}$ be the intensity map (represented with a \SI[number-unit-product = -]{8}{\bit} grayscale image) of a texture frame and $I_{\text{depth}}$ be the corresponding depth map (represented with another \SI[number-unit-product = -]{8}{\bit} grayscale image) in an MVD sequence. Let $\hat{I}$ be the motion-compensated predicted image, $\Delta I = I - \hat{I}$ be the corresponding residual frame of signed integers, and $\Delta \mathbb{I}$ be the corresponding absolute residual frame of residual magnitudes for a grayscale image $I$.

\begin{figure}[!tb]
	\centering	
	
	\subfloat[]{\includegraphics[width=0.48\columnwidth]{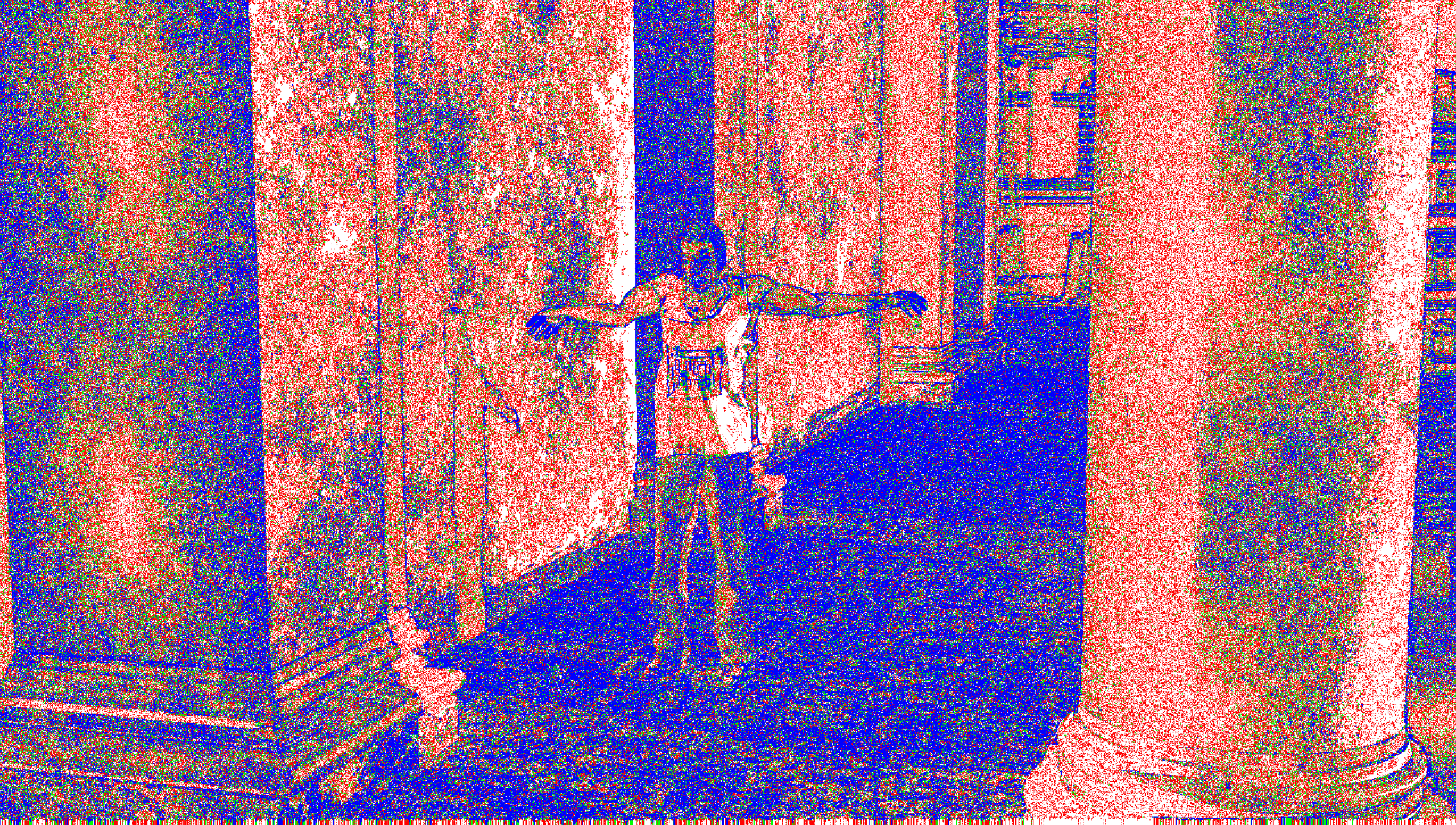}}
	\hfill	
	\subfloat[]{\includegraphics[width=0.48\columnwidth]{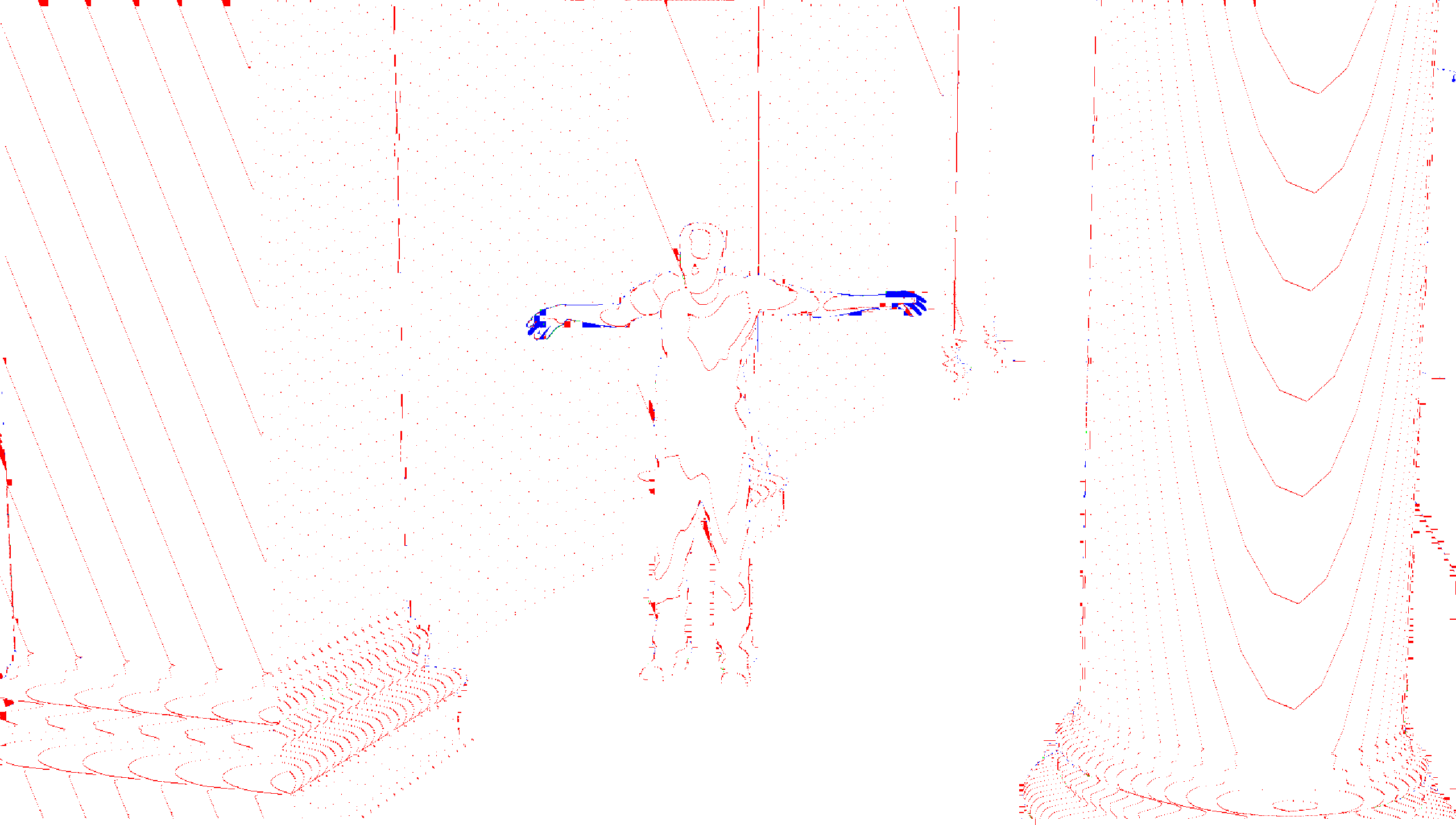}}
	
	\subfloat[]{\includegraphics[width=0.48\columnwidth]{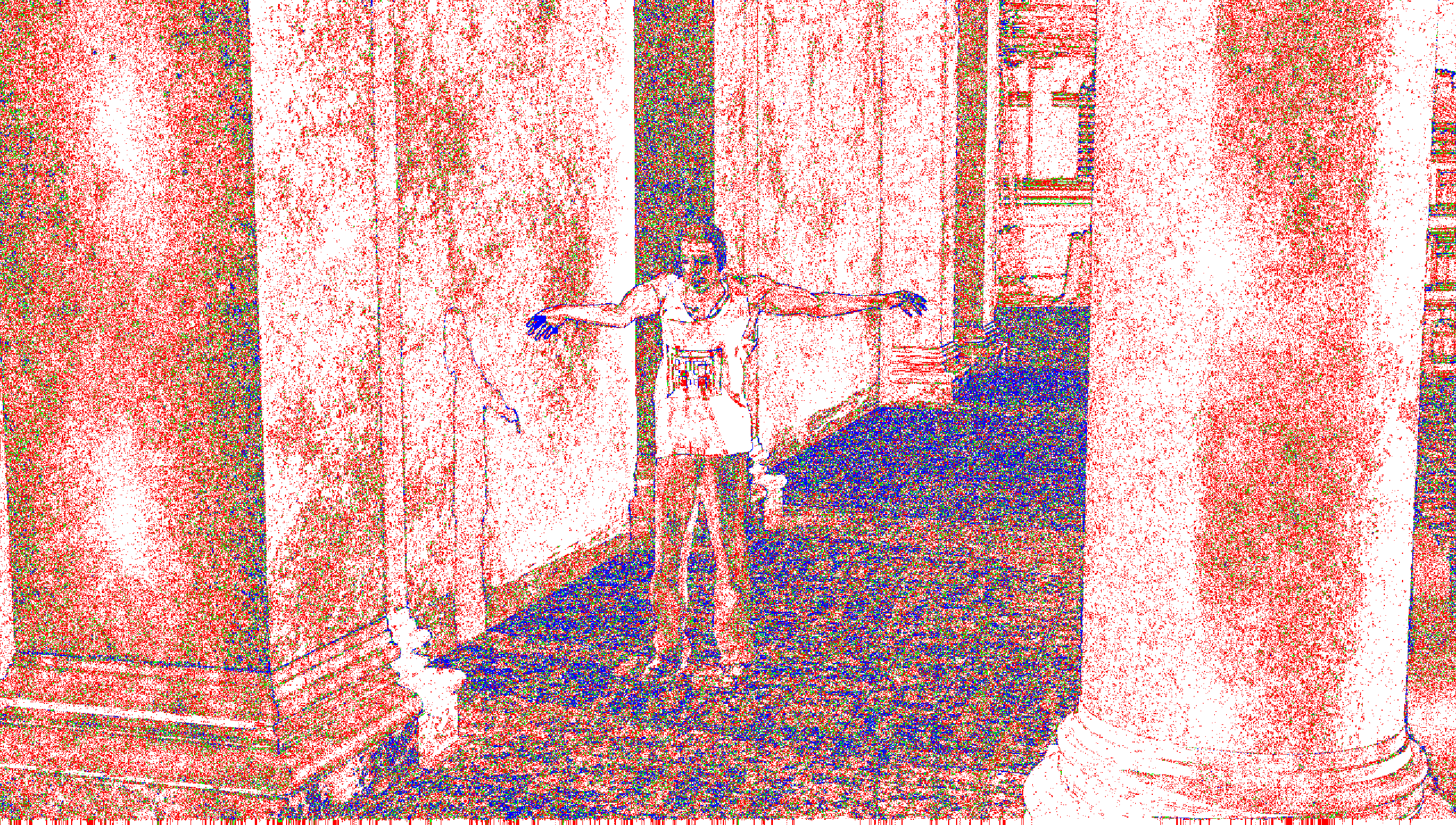}}
	\hfill	
	\subfloat[]{\includegraphics[width=0.48\columnwidth]{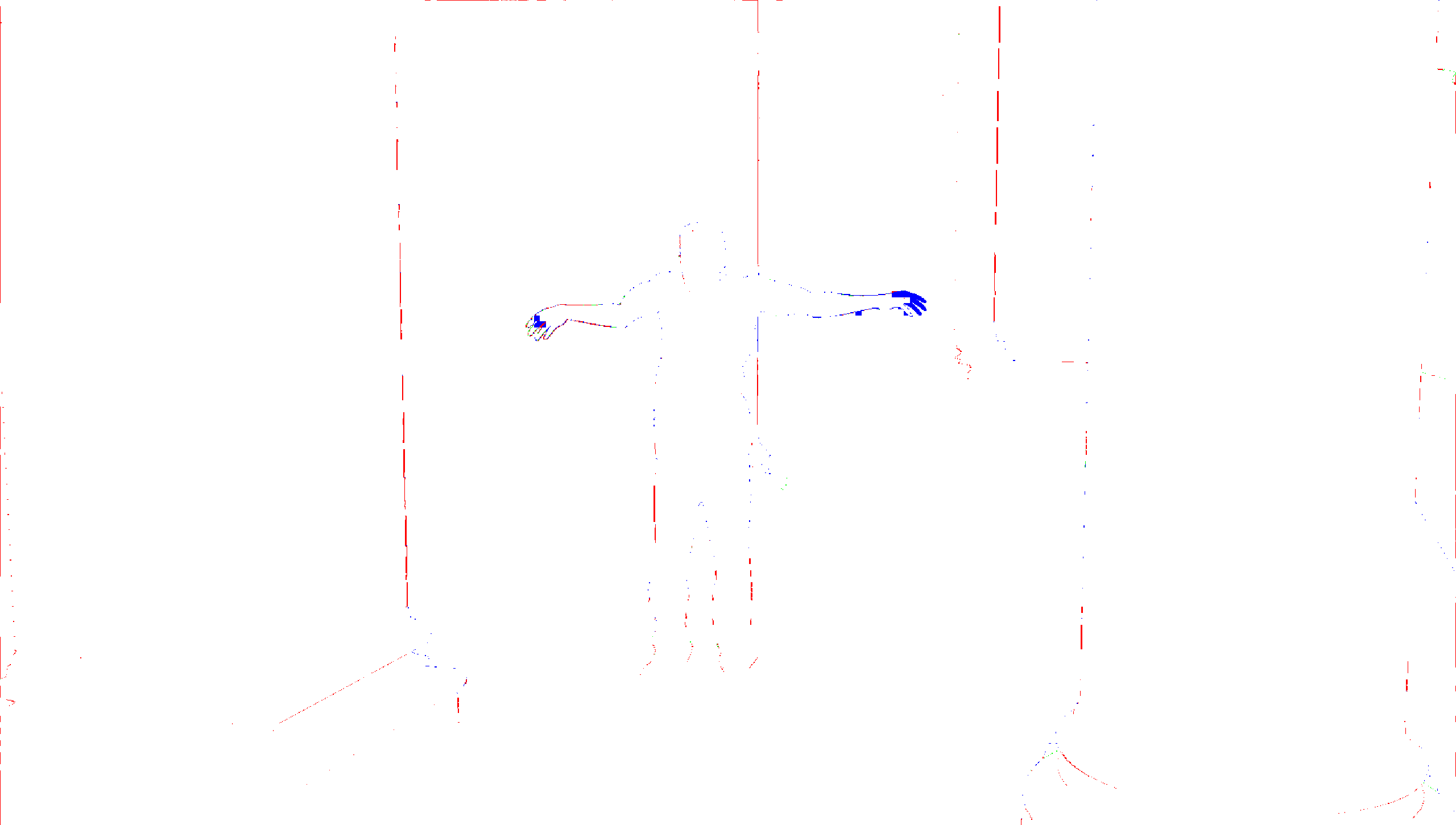}}
	
	\caption{Magnitude of residuals for (a) lossless intensity; (b) lossless depth; (c) quantised intensity; and (d) quantised depth maps of \textit{UndoDancer} sequence (view~1, frame~23), where white, red, green, and blue represent magnitudes \numlist{0;1;2;\geq3}, respectively, and $Q = 3$ is used for quantisation. (best viewed in colour)}
	\label{fig:Ch3_resstattex}
\end{figure}

\figurename\,\ref{fig:Ch3_resstattex} presents typical $\Delta \mathbb{I}_{\text{texture}}$ and $\Delta \mathbb{I}_{\text{depth}}$, both lossless and with spatial-domain quantisation by scalar step $Q = 3$, for an arbitrary frame~23 in view~1 of \textit{UndoDancer} multiview standard test sequence. The following characteristically unique observations can be made on clustering tendency in depth residuals.

Firstly, due to high spatio-temporal correlations in depth maps, almost \SI{90}{\percent} values in $\Delta \mathbb{I}_{\text{depth}}$ are zeros, compared to only \SIrange{20}{30}{\percent} in $\Delta \mathbb{I}_{\text{texture}}$. Secondly, dominant zeros are clustered densely in depth residuals; whereas non-dominant zeros are spread thinly across texture residuals. Thirdly, when quantisation is applied, the clustering tendency in depth residuals is intensified. Although, zeros become dominant (around \SI{60}{\percent}) in texture residuals, the clustering tendency is still not that prominent. Only when the quantisation level is increased, clustering in smaller values are introduced for the high motion sequence. Finally, the dynamic range in $\Delta \mathbb{I}_{\text{depth}}$ is significantly low, compared to $\Delta \mathbb{I}_{\text{texture}}$. In both lossless and quantised domains, the dynamic range in the former is \numrange[range-phrase = --]{30}{40} lower. 

Similarly strong clustering tendency is evident in CTU divisions (\figurename\,\ref{fig:ch3_figexample10c}), CU prediction modes (\figurename\,\ref{fig:ch3_figexample10c}), and motion components (\figurename\,\ref{fig:ch3_figexample10a}) of a depth map. Such high clustering tendency in depth residual frames leads to localised regions of skewed probability distribution. This may be exploited by arithmetic coding, along with the low dynamic range, to achieve significant compression efficiency after isolating the regions with any density-based partitioning.

\section{Proposed Depth Map Sequence Coder}
\label{sec:Ch3_depthcoder}

In this section, an independent depth map sequence coder is proposed based on the above concept. It decides quad-tree based CTU divisions (similar to 3D-HEVC) and the prediction modes of CUs (of size \SIlist[product-units=single]{64 x 64;32 x 32;16 x 16;8 x 8}{pixel}) for a particular frame based on a rate-distortion optimisation scheme (Section~\ref{sec:ch3_ctudivisiontechnique}). The CU-level signed motion-compensated depth residuals are first quantised, if near-lossless mode is used, and then mapped to unsigned integers using a generic ranking-based residual mapping technique (Section~\ref{sec:ch3_ranking}). Depth syntax and mapped-residual data are  arranged into a number of frame-level bitmaps or integer maps (intmaps) (Section~\ref{sec:ch3_mapformation}). The clustering-correlations in these maps are exploited using BTBD partitioning (Section~\ref{sec:Ch3_btbd}), where large homogeneous blocks with similar values are isolated and then encoded using context-adaptive arithmetic coding. The leaf nodes generated from map partitioning are used as the coding blocks. Unlike 3D-HEVC, coding block size is independent of CU size and adaptive to clustering-correlations. Finally, these coding blocks are encoded using CAAC (Section~\ref{sec:Ch3_encoding}).

\subsection{CTU division and prediction mode selection}
\label{sec:ch3_ctudivisiontechnique}

Let $B(I,r,c,m)$ denote a CU of size $m \times m$\,\si{pixel} ($m=2^{6-k}$ and $0 \leq k \leq 3$) in depth frame $I$ at pixel coordinate $(r,c)$, i.e., the CU represents $I(r,\ldots,r+m-1;c,\ldots,c+m-1)$ values of the frame covering all pixel coordinates $(i,j)$'s such that $i \in \{r,\ldots,r+m-1\}$ and $j\in \{c,\ldots,c+m-1\}$. Each frame $I$ of size $H \times W$\,\si{pixel} is divided into $\frac{H}{m}$ rows and $\frac{W}{m}$ columns of non-overlapping coding units such that the CU at row $r_{m}$ and column $c_{m}$ refers to $B(I,(r_{m}-1)m+1,(c_{m}-1)m+1,m)$, for all $1 \le r_{m} \le \frac{H}{m}$ and $1 \le c_{m} \le \frac{W}{m}$.

In inter prediction coding, each of the $\frac{H}{m} \times \frac{W}{m}$ CUs in the current frame $I_{t}$ is predicted from the already encoded reference frame $I_{t-1}$ using block-based motion search. Let $MV_{m}(p,r_m,c_m)$ denote the motion vector (MV) components of CU at row $r_{m}$ and column $c_{m}$ where $p\in \{1~\equiv~x\text{-component},2~\equiv~y\text{-component}\}$. For intra prediction coding, each of the $\frac{H}{m} \times \frac{W}{m}$ CUs in the current frame $I_{t}$ is predicted from already coded similar neighbourhood CUs in $I_{t}$ (using CALIC's gradient adjusted predictor~\cite{calic}).

For a fixed coding unit size $m \times m$\,\si{pixel}, let $\hat I_{t_{m}}^{A}$ and $\Delta I_{t_{m}}^{A}=I_{t}-\hat I_{t_{m}}^{A}$ denote the predicted and residual frame, respectively, such that for each CU $B(I,(r_{m}-1)m+1,(c_{m}-1)m+1,m)$, the co-located blocks $B(\hat I_{t_{m}}^{A},(r_{m}-1)m+1,(c_{m}-1)m+1,m)$ and $B(\Delta I_{t_{m}}^{A},(r_{m}-1)m+1,(c_{m}-1)m+1,m)$ represent its prediction and residual, respectively, for all $1 \le r_{m} \le \frac{H}{m}$, $1 \le c_{m} \le \frac{W}{m}$, and $A \in \{\text{inter},\text{intra}\}$.

For each $64 \times 64$\,\si{pixel} CTUs in frame $I_{t}$, division and mode selection decisions are made based on RDO. First, the optimal prediction mode of the CTU is decided and then, a quad-tree based sub-division into four CUs is recursively decided if further optimality can be achieved. Compared to 3D-HEVC, the RDO of the proposed coder is simple. The maximum pixel-level distortion $D$ in a frame is controlled using a fixed scalar quantisation step $Q = 2D + 1$. While operating at near-lossless mode ($Q > 1$), distortion depends solely on the scalar quantisation step $Q$, irrespective of prediction modes. Hence, the bitrate-optimal prediction mode for any CU $B(I,(r_{m}-1)m+1,(c_{m}-1)m+1,m)$ is decided by comparing the \emph{estimated} bits needed to encode $B(\Delta I_{t_{m}}^{\textrm{intra}},(r_{m}-1)m+1,(c_{m}-1)m+1,m)$ against the same for $B(\Delta I_{t_{m}}^{\textrm{inter}},(r_{m}-1)m+1,(c_{m}-1)m+1,m)$ and the MV $(MV_{m}(1,r_{m},c_{m}),MV_{m}(2,r_{m},c_{m}))$. For $m > 2^3$, the CU is recursively divided into four $\frac{m}{2} \times \frac{m}{2}$\,\si{pixel} CUs only if bits are saved. A code-length estimator (Section~\ref{sec:code-length}) for encoding the residual block with CAAC, after mapping to unsigned integers (Section~\ref{sec:ch3_ranking}), is used. Bits of encoding each MV component is estimated $2\lceil \log_{2}(|MV_{m}( p, r_{m} , c_{m})|+1)\rceil+1$\,\si{\bit}, with signed Exp-Golomb codes~\cite{Paper15_richardson2011}, $p \in \{1,2\}$.

\begin{figure}[!tb]
	\centering
	\includegraphics[width=1\columnwidth]{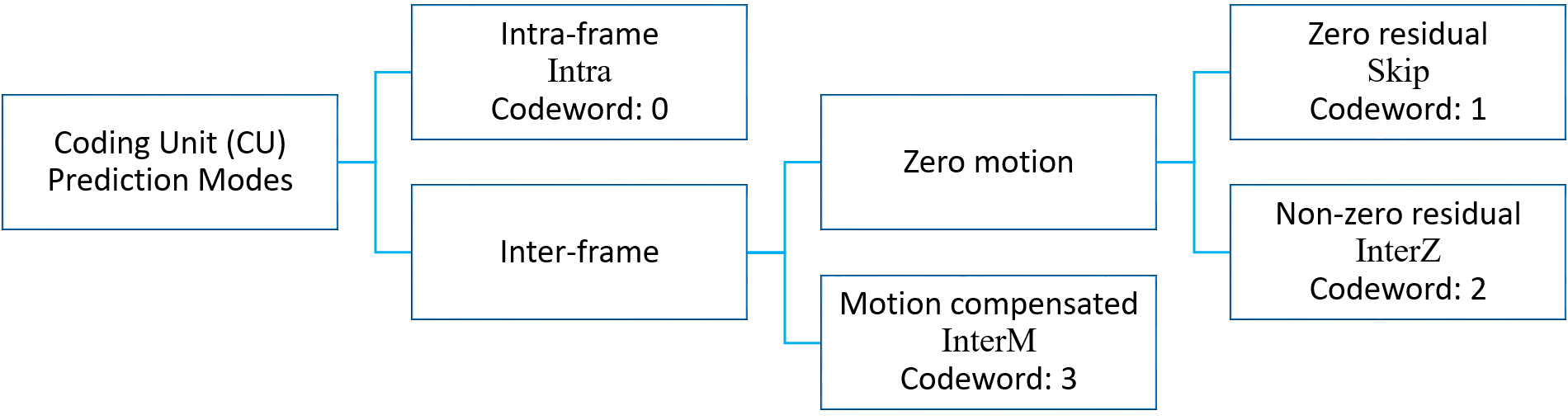}
	\caption{Prediction modes of depth map coding units.}
	\label{fig:ch3_figcodingMB}
\end{figure}

Four different predictive modes are used as depicted in \figurename\,\ref{fig:ch3_figcodingMB} with their hierarchical relationships. A CU is intra- or inter-predicted. If intra-predicted (\si{Intra}), only the residuals need to be encoded. Otherwise, there are two possibilities: (a) both MV components are zeros (zero motion), when the best match is found in the temporally co-located CU; or (b) at least one MV component is non-zero. In the second case (\si{InterM}), residuals as well as the MV need to be encoded for the motion-compensated CU. In the first case, there can be further two possibilities: (a) CU exactly matches the co-located CU (\si{Skip}) for which neither residual nor MV need encoding; or (b) otherwise (\si{InterZ}) for which only the residuals need to be encoded.

\subsection{Residual mapping with quantisation}
\label{sec:ch3_ranking}

Consider predictive coding of a sequence $X = x_0,x_1,\ldots$ drawn from an integer-valued source. Let $\hat x_{t}$ be the integer-valued prediction for $x_t$, computed from the past samples
$x_{0}, x_{1}, \ldots, x_{t-1}$, and $\epsilon_t = x_t -\hat x_t$ be its signed residual.

With an 8-bit grayscale image source, the domains of $x$ and $\epsilon$ are $[0,R]$ and $[-R,R]$, respectively, where $R = 2^8 - 1$. For any integer-valued prediction $\hat x$, however, there can be at most $R + 1$ distinct signed residual values, $V_{\hat x} = \{-\hat x,1 - \hat x,\ldots,R - \hat x\}$. It is, therefore, possible to establish a one-to-one correspondence (aka bijection) map between the signed residuals and the unsigned integer domain $[0,R]$ by using the rank $r_\epsilon$ of a signed residual $\epsilon$ in the corresponding $V_{\hat x}$ w.r.t. the magnitude as follows:
\begin{equation}
r_\epsilon = \begin{cases}
2\epsilon, & |\epsilon| \le \min(\hat x,R - \hat x) \wedge \epsilon \ge 0; \\
-2\epsilon - 1, & |\epsilon| \le \min(\hat x,R - \hat x) \wedge \epsilon < 0;\\
\min(\hat x,R - \hat x) + |\epsilon|, & |\epsilon| > \min(\hat x,R - \hat x).
\end{cases}
\label{eq:rankmap}
\end{equation}

This ranking based mapping is partly influenced by the Rice mapping~\cite{Rice-Golomb}. Similar to the Rice mapping, starting from zero, the ranking technique maps successive -ve and +ve residuals to odd and even values, respectively, until one side is exhausted, then it uses linear mapping for the other side. In doing so, it keeps the dynamic range unchanged, whereas the Rice mapping can potentially double the dynamic range and hence degrade compression efficiency of arithmetic coding.

Spatial-domain quantisation guarantees that every reconstructed sample does not diverge from the original signal by more than a preset amount $D \ge 0$. The classical approach to near-lossless coding is to uniformly quantise the integer-valued residual $\epsilon$ into quantisation bins of size $Q=2D+1$, with reproduction at the centre of the bins. Let $\epsilon_{Q}=\lceil \frac{\epsilon}{Q} \rfloor$ be the quantised-residual rounded to the nearest integer (denoted by the $\lceil \cdot \rfloor$ operator), with scalar quantisation step $Q$. Note that the inverse-quantised value $Q \epsilon_{Q}$ has quantisation noise $\epsilon - Q \epsilon_{Q}$ in the range $[-D,D]$ for all $Q > 1$. No quantisation takes place for $Q = 1$, which is used for coding at lossless mode.

The ranking based mapping can be easily extended to quantised-residuals. Note that the quantisation process effectively reduces the domain of residuals to $[-\lceil \frac{R}{Q} \rfloor,\lceil \frac{R}{Q} \rfloor]$. Hence, the rank $r_{\epsilon_Q}$ of $\epsilon_{Q}$ in the corresponding $V_{\lceil \frac{\hat x}{Q} \rfloor}$ w.r.t. the magnitude can be estimated by substituting $\epsilon$, $\hat x$, and $R$ in (\ref{eq:rankmap}) with $\epsilon_Q$, $\lceil \frac{\hat x}{Q} \rfloor$, and $\lceil \frac{R}{Q} \rfloor$, respectively.

\subsection{Data map formation}
\label{sec:ch3_mapformation}

To exploit strong clustering tendency in depth maps, CTU divisions, CU prediction modes, MV components, and quantised-residual ranks of each frame $I_t$ are arranged into a number of bitmaps/intmaps so that BTBD partitioning can be applied to isolate clusters before encoding.

\subsubsection{CTU division bitmaps}

Quad-tree based CTU division decisions of a frame is represented jointly with three bitmaps. For each of the first three division levels $l \in \{0,1,2\}$, where decisions on splitting a $2^{6 - l} \times 2^{6 - l}$\,\si{pixel} CU into four $2^{5 - l} \times 2^{5 - l}$\,\si{pixel} CUs are made, a bitmap $\mathcal{M}_{\text{div}_{2^{6-l}}}$ of size $\frac{H}{2^{6-l}}\times \frac{W}{2^{6-l}}$ is used, where each bit represents whether the corresponding $2^{6 - l} \times 2^{6 - l}$\,\si{pixel} CU is divided further (\num{1}) or not (\num{0}). Note that no bitmap is needed for \SI[product-units=single]{8 x 8}{pixel} CUs as these are never split.

For the two lower levels, $l \in \{1,2\}$, bits corresponding to non-existent CUs, due to not-split decisions at a level above, are ignored. A virtual don't-care symbol $\dc$ is used to identify the ignored CUs by setting $\mathcal{M}_{\text{div}_{2^{6-l}}}(r_{2^{6-l}},c_{2^{6-l}}) = \dc$ if $\exists k < l: \mathcal{M}_{\text{div}_{2^{6-k}}}( \lceil \frac{r_{2^{6-l}}}{2^{l - k}} \rceil,\lceil \frac{c_{2^{6-l}}}{2^{l - k}} \rceil ) = 0$. The symbol $\dc$ is considered virtual as it can always be deducted from other maps and it is never encoded.

In order to deal uniformly with variable CU sizes, a frame is considered as a $\frac{H}{8}\times \frac{W}{8}$ grid of non-overlapping CUs of the smallest size and a virtual significance bitmap $\mathcal{M}_{\dc}$ of size $\frac{H}{8}\times \frac{W}{8}$ is used. The CTU at row $r_{64}$ and column $c_{64}$ is represented by the block of \SI[product-units=single]{8 x 8}{\bit} in $\mathcal{M}_{\dc}$ starting at coordinate $(8 r_{64} - 7,8 c_{64} - 7)$. In that CTU, for each leaf (no further split) CU of size $2^{6-k} \times 2^{6-k}$\,\si{pixel}, $k \in \{0,1,2,3\}$, only the top-left bit position of the corresponding $2^{3-k}\times 2^{3-k}$\,\si{\bit} in $\mathcal{M}_{\dc}$ is set to \num{0} and the remaining bits are set to \num{1}. Bitmap $\mathcal{M}_{\dc}$ is considered virtual as it is never encoded; it just helps simplifying the process of defining the remaining data maps.

\subsubsection{Mode intmap}

Four CU prediction modes \si{Intra}, \si{Skip}, \si{InterZ}, and \si{InterM}  are assigned code word \numlist{0;1;2;3}, respectively. Prediction modes of all leaf CUs in a frame are represented with an intmap $\mathcal{M}_{\text{mode}}$ of size $\frac{H}{8} \times \frac{W}{8}$. For each bit $0 \in \mathcal{M}_{\dc}$, the co-located integer in $\mathcal{M}_{\text{mode}}$ is set to the code word of the prediction mode used by the corresponding leaf CU. The remaining coordinates of  $\mathcal{M}_{\text{mode}}$ are ignored, i.e., $\mathcal{M}_{\text{mode}}(r_8,c_8) = \dc$ if $\mathcal{M}_{\dc}(r_8,c_8) = 1$.

\subsubsection{Zero/non-zero MV component bitmap}

Depth MV components are predominantly zero-valued. For any leaf CU using a prediction mode other than \si{InterM}, the MV either does not exist (\si{Intra}) or it is fixed (\si{Skip} and \si{InterZ}) with both components zero-valued. There is no need to encode MV for these CUs.

A 3D bitmap $\mathcal{M}_{\overline {\text{MVZ}}}$ of size $2 \times \frac{H}{8}\times \frac{W}{8}$ is used to represent the zero/non-zero binary-classification of all encoded MV components in a frame. For each bit $0 \in \mathcal{M}_{\dc}$, the co-located 2-bits in $\mathcal{M}_{\overline{\text{MVZ}}}$ are set to $0$ (zero-valued) or $1$ (non-zero-valued), based on two MV components of the corresponding leaf CU, if that CU is using `Inter-MC' prediction mode. The remaining coordinates of  $\mathcal{M}_{\overline {\text{MVZ}}}$ are ignored, i.e., $\mathcal{M}_{\overline {\text{MVZ}}}(p,r_8,c_8) = \dc$ if $\mathcal{M}_{\dc}(r_8,c_8) = 1$ or $\mathcal{M}_{\text{mode}}(r_8,c_8) \not= 3$ for all $p \in \{1,2\}$.

\subsubsection{Residual intmap}

For any leaf CU using prediction mode `Skip', the residuals are fixed, all zero-valued, even after quantisation and ranking based mapping. There is no need to encode the residuals for these CUs.

An intmap $\mathcal{M}_{\text{res}}$ of size $H\times W$ is used to represent the pixel-level ranks of quantised-residuals that need encoding. For each bit $0 \in \mathcal{M}_{\dc}(r_8,c_8)$, ranks of quantised-residuals of the corresponding leaf CU are stored in $\mathcal{M}_{\text{res}}$ starting at coordinate $(8r_8 - 7,8c_8 - 7)$ if that CU is not using `Skip' prediction mode. Otherwise, these coordinates in $\mathcal{M}_{\text{res}}$ are ignored.

\subsection{Map partitioning with BTBD}
\label{sec:Ch3_btbd}

The BTBD map partitioning uses a greedy heuristic on minimizing the estimated code-length of encoding partitions with CAAC.

\subsubsection{Context modelling}
\label{sec:context}

Context $\mathbb{C}_{\mathcal{M}}(\cdot)$ of any coordinate $(\cdot)$ in a data map $\mathcal{M}$ is calculated using the values in adjacent coordinates along all dimensions. If any adjacent value is the don't-care symbol $\dc$, it is replaced with \num{0}; except for $\mathcal{M}_{\text{mode}}$ maps, where $\dc$ is replaced with the code word of the corresponding leaf CU. For $\mathcal{M}_{\text{res}}$, the magnitude of the inverse-quantised residual $|Q\epsilon_Q|$ corresponding to the adjacent value $r_{\epsilon_Q}$ is used. Let $\mathcal{A}_d(\mathcal{M}(\cdot))$ denote the adjacent value along dimension $d$ of  coordinate $(\cdot)$ in $\mathcal{M}$ as defined above.
     
Among the six data maps $\mathcal{M}_{\text{div}_{64}}$, $\mathcal{M}_{\text{div}_{32}}$, $\mathcal{M}_{\text{div}_{16}}$, $\mathcal{M}_{\text{mode}}$, $\mathcal{M}_{\overline {\text{MVZ}}}$, and $\mathcal{M}_{\text{res}}$, the first five represent nominal (non-collatable) data and only the last one represents ordinal (collatable) data.

For a nominal map $\mathcal{M}$ of dimension $\mathcal{D} \in \{2,3\}$, all possible combinations of adjacent values are considered as different contexts that are mapped to unique integer values as
\begin{equation}
\mathbb{C}_{\mathcal{M}}(\cdot) = \sum_{d = 1}^\mathcal{D} n^{d - 1} \mathcal{A}_d(\mathcal{M}(\cdot)),  
\end{equation}
where $n = 2 \text{ and } 4$ for bitmap and intmap, respectively. With nominal bitmaps, the domain of $\mathbb{C}_{\mathcal{M}}$ is $[0,2^2 - 1]$ and $[0,2^3 - 1]$ for 2D maps ($\mathcal{M}_{\text{div}_{64}}$, $\mathcal{M}_{\text{div}_{32}}$, and $\mathcal{M}_{\text{div}_{16}}$) and 3D map ($\mathcal{M}_{\overline {\text{MVZ}}}$), respectively. The same for the nominal intmap $\mathcal{M}_{\text{mode}}$ is $[0,4^2 - 1]$.

For the ordinal map $\mathcal{M}_{\text{res}}$, the range of an adjacent value (magnitude of inverse-quantised residual) is $[0,255]$. Collectively, the adjacent values along two dimensions can have $2^{16}$ combinations that are far too many to consider as different contexts. Nevertheless, the ordinality of residuals can be effectively used to define only four contexts by dividing the range $[0,510]$ of the sum of adjacent residuals into four bins of geometrically increasing length as follows:
\begin{equation}
\mathbb{C}_{\mathcal{M}_{\text{res}}}(\cdot)= \begin{cases}
0, & 0\le \mathcal{A}_1(\mathcal{M}(\cdot)) + \mathcal{A}_2(\mathcal{M}(\cdot)) \le 4;\\
1,&  5\le \mathcal{A}_1(\mathcal{M}(\cdot)) + \mathcal{A}_2(\mathcal{M}(\cdot)) \le 22;\\
2, & 23\le \mathcal{A}_1(\mathcal{M}(\cdot)) + \mathcal{A}_2(\mathcal{M}(\cdot)) \le 117;\\
3, & 118\le \mathcal{A}_1(\mathcal{M}(\cdot)) + \mathcal{A}_2(\mathcal{M}(\cdot)) \le 510,
\end{cases}
\label{eq:ch3_contextres2}
\end{equation}    
Note that the specific bin edges in (\ref{eq:ch3_contextres2}) are empirically obtained through sensitivity analysis on multiview test sequences.

\subsubsection{Code-length estimation of CAAC}
\label{sec:code-length}

Let $\mathcal{M}$ be a bitmap/intmap having $N$ non-ignored valued, drawn from the range $[0,R]$, where context-model $\mathbb{C}_{\mathcal{M}}$ has been used to divide them into $k$ subsets $\mathbb{C}_{\mathcal{M},i}$'s, $0\le i < k$. Let $n_{i,j}$ be the count of value  $j$ in context $\mathbb{C}_{\mathcal{M},i}$ such that $N_i = \sum_{j=0}^{R}n_{i,j}$ and $\sum_{i=0}^{k-1}N_{i} =N$. The code-length $\mathcal{L}(\mathcal{M})$ of encoding $\mathcal{M}$ with CAAC can be estimated by adding the zero-order entropy $\mathcal{H}_{0} (\mathbb{C}_{\mathcal{M},i})$ and the associated model cost $\mu(\mathbb{C}_{\mathcal{M},i})$ of all $k$ context as follows:
\begin{equation}
\begin{split}
\mathcal{L}(\mathcal{M}) & = \left \lceil \sum_{i=0}^{k-1} N_{i}\mathcal{H}_{0} (\mathbb{C}_{\mathcal{M},i})+ \mu(\mathbb{C}_{\mathcal{M},i}) \right \rceil\\
& =\left \lceil - \sum_{i=0}^{k-1} \sum_{j=0}^{R} n_{i,j} \log_{2} p_{i,j} + \mu(\mathbb{C}_{\mathcal{M},i})\right \rceil \si{\bit},
\end{split}
\label{eq:ch3_codelength1}
\end{equation}
where $p_{i,j} = \frac{n_{i,j}}{N_{i}}$ is the static probability of symbol $j$ in context $i$.

Model cost of arithmetic coding is the implicit overhead of encoding probability model parameters $p_{i,j}$'s with sufficient precision so that correct decoding is guaranteed. For sufficiently large $N_i$, it has been shown that each free (independent) parameter incurs $\frac{1}{2} \log_{2} N_i$ bits of model cost~\cite{491334}. Considering the density constraint $\sum_{j = 0}^R p_{i,j} = 1$, each context $\mathbb{C}_{\mathcal{M},i}$ has at most $R$ free parameters and hence,
\begin{equation}
\label{eq:ch_back_modelcosteq1}
\mu(\mathbb{C}_{\mathcal{M},i}) \approx \frac{R}{2} \log_{2} N_i  \,\si{bit},
\end{equation}
for all $0 \le i < k$.

When $N_i$ is not sufficiently large, the number of free parameters is estimated more accurately using the following empirically obtained correction:
\begin{equation}
\label{eq:ch_back_modelcosteq2}
\hat{R}= \frac{R}{2^{\frac{R + 1}{\log_{2} N_i} \left( 2^{\frac{1-\log_{2} (R + 1)}{2}} - \hat{p}_i \right)}},
\end{equation}
where $\hat{p}_i$ is the estimated parameter of the geometric probability density function that fits the distribution of residuals in $\mathbb{C}_{\mathcal{M},i}$ best. Parameter $\hat{p}_i$ is estimated from $n_{i,j}$'s by the method of moments~\cite{ Kapadia1975} as
\begin{equation}
\label{eq:ch_back_modelcosteq3}
\hat{p}_i=\frac{(R+2)\sum_{j=0}^{R}n_{i,j}-2\sum_{j=0}^{R}j n_{i,j}}{(R+1)\sum_{j=0}^{R} j n_{i,j}-\sum_{j=0}^{R}j^2 n_{i,j}}.
\end{equation}

\begin{figure}[!tb]
	\centering
	
	\subfloat{\epsfig{figure=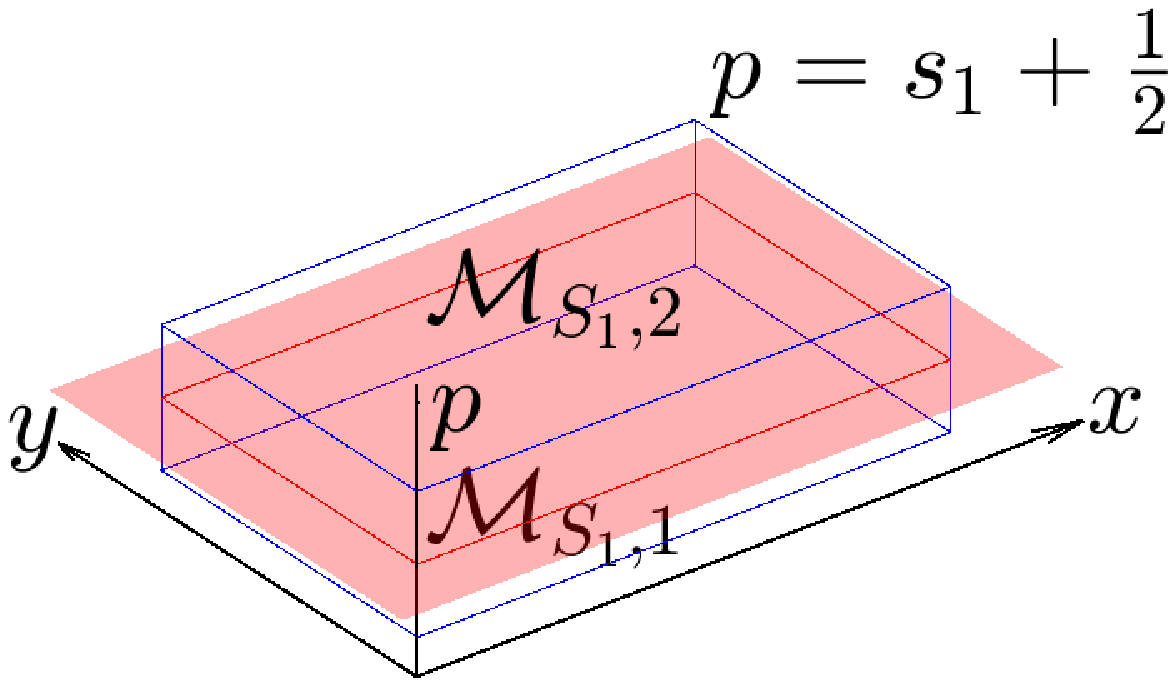,width=0.33\columnwidth,trim={2cm 1.5cm 0.75cm 2cm},clip}}		
	\hfill
	\subfloat{\epsfig{figure=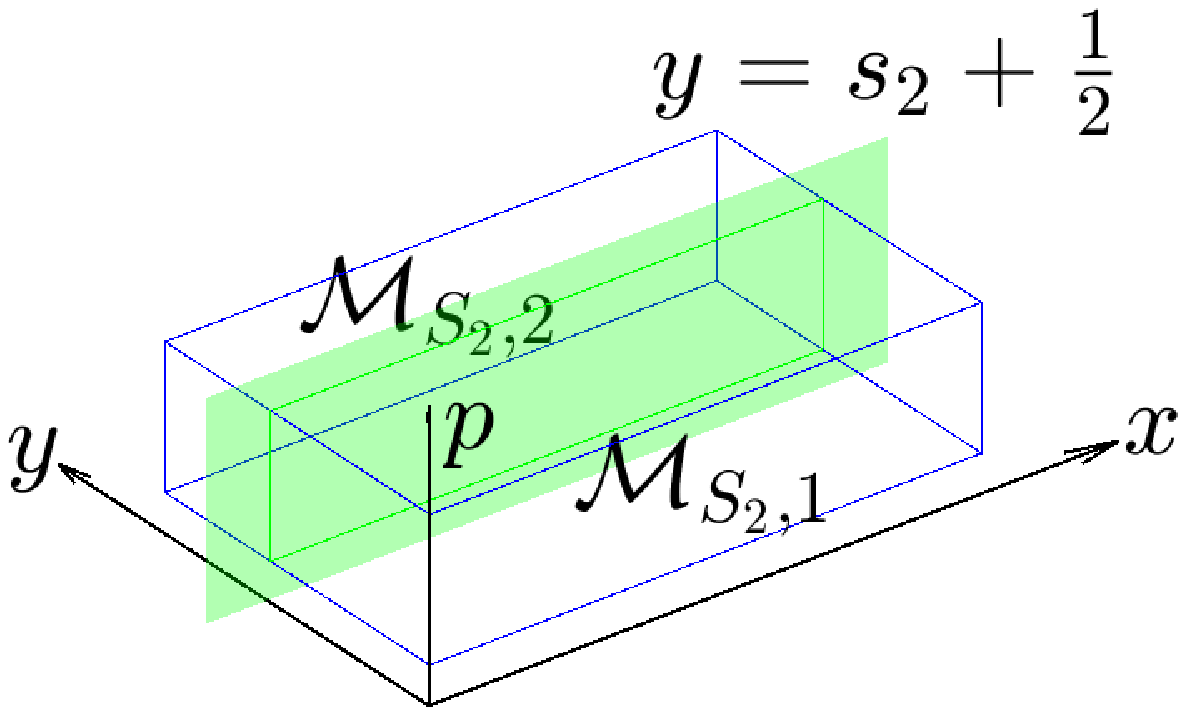,width=0.33\columnwidth,trim={2cm 1.5cm 0.75cm 2cm},clip}}		
	\hfill
	\subfloat{\epsfig{figure=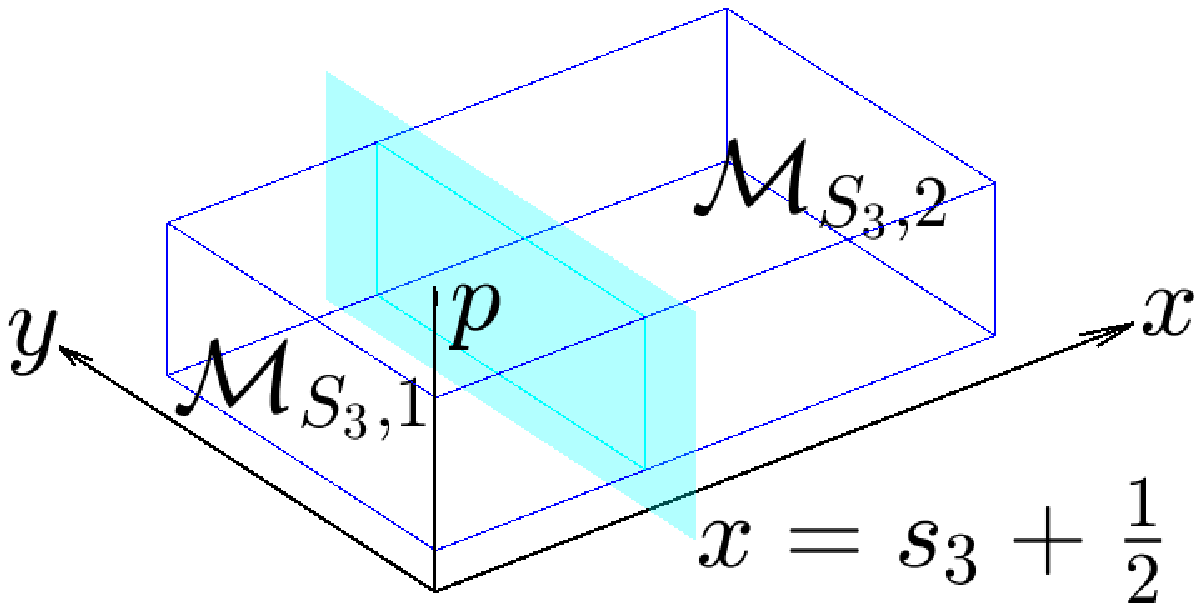,width=0.33\columnwidth,trim={1.5cm 1.5cm 0.75cm 2cm},clip}}		
	
	\caption{Splitting a 3D map into halves with a plane orthogonal to (left) $p$-axis; (middle) $y$-axis; and (right) $x$-axis. (best viewed in colour)}
	\label{fig:ch2_figdiv}
\end{figure}

\subsubsection{Binary tree based decomposition (BTBD)}
\label{sec:BTBD}

BTBD uses a greedy divide-and-conquer heuristic. The {\em divide} step tries to optimally split a bitmap/intmap into halves recursively so that encoding bits are saved. Consider a map $\mathcal{M}$ of size $\psi_1 \times \psi_2 \times \psi_3$ (representing its length in $p$-, $y$-, and $x$-axis, respectively) with $N \le \prod_{d = 1}^{3} \psi_d$ non-ignored elements. If $\mathcal{M}$ contains only one symbol, it is classified as a Type~$\RN{1}$ (all $0$'s) or Type~$\RN{2}$ (all $1$'s for bitmap and all same non-zero (SNZ) values for intmap) leaf node. Otherwise, it is split recursively into two halves with a plane orthogonal to $p$-, $y$-, or $x$-axis (see \figurename\,\ref{fig:ch2_figdiv}), respectively classified as Type P, Type Y, or Type X split, such that the overall code-length is minimised.

Let a split $s_d$ of $\mathcal{M}$ along dimension $d \in \{1,2,3\}$ form two cuboid halves $\mathcal{M}_{s_{d},1}$ and $\mathcal{M}_{s_{d},2}$. If these cuboids are encoded independently, the overall code-length $\mathcal{L}_{s_{d}}$ due to split $s_d$ can be measured as
\begin{equation}
\mathcal{L}_{s_{d}}(\mathcal M)=\mathcal{L}(\mathcal{M}_{s_{d},1}) + \mathcal{L}(\mathcal{M}_{s_{d},1}) + \mu_s(s_d),
\label{eq:entropyHall}
\end{equation}
where $\mu_s(s_d)$ is the split cost, representing the overhead of encoding the split dimension and position. Split dimension $d$ is encoded with variable-length Huffman codes (see Table\,\ref{tab:Ch3_huffmantreenode}).  A split $s_d$ actually represents a plane $p = s_1 + \frac{1}{2}$, $y = s_2 + \frac{1}{2}$, and $x = s_3 + \frac{1}{2}$ for $d \in \{1,2,3\}$, respectively. Along each dimension $d$, there are $\psi_d(\mathcal{M}) - 1$ possible splits denoted by the range $1 \le s_d < \psi_d(\mathcal{M})$. For bitmaps, all possible splits are considered and hence, $s_d$ is encoded with fixed-length codes of $\log_2 (\psi_d(\mathcal{M}) - 1)$ bits. For integer maps, only the half-way split $s_d = \lceil \frac{1}{2} (\psi_d(\mathcal{M}) - 1) \rceil$ along each dimension $d$ is considered and hence, $s_d$ requires no encoding.

The greedy heuristic finds the optimal split $s^*_{d^*}$ among all possible splits under consideration such that $\mathcal{L}_{s_{d}}(\mathcal M)$ is minimised as follows:
\begin{align}
s^*_d &= \begin{dcases} \argmin_{1 \le s_{d} < \psi_d(\mathcal{M})}\mathcal{L}_{s_{d}}(\mathcal{M}), & \text{$\mathcal{M}$ is a bitmap}; \\
\left\lceil (\psi_d(\mathcal{M}) - 1)/2 \right\rceil, & \text{$\mathcal{M}$ is an intmap}. \end{dcases} \\
d^* &= \argmin_{d \in \{1,2,3\}}\mathcal{L}_{s^*_d}(\mathcal{M}).
\label{eq:ch2_optimalsp1}
\end{align}

\begin{algorithm}[!tb]
	\caption{BTBD ($\mathcal{M}$)}
	\begin{algorithmic}[1]
		\REQUIRE Bitmap $\mathcal{M}$\\
		\ENSURE Binary partition-tree $T$\\
		\IF {$\mathcal{M}$ contains only one symbol (excluding $\dc$)}
		\STATE{$T \gets \langle \text{Type}~\RN{1} \text{ or}~\RN{2} \text{ leaf node} \rangle$;}
		\ELSE
		\STATE{Find the optimal split $s^*_{d^*}$;}
		\STATE{$T_1 \gets \text{BTBD}(\mathcal{M}_{s^*_{d^*},1})$;} 
		\STATE{$T_2 \gets \text{BTBD}(\mathcal{M}_{s^*_{d^*},2})$;}
		\IF {$\mathcal{L}_{s^*_{d^*}}(\mathcal{M}) < \mathcal{L}(\mathcal{M})$}
		\STATE{$T \gets \langle \text{Type } d^* \text{ internal node},T_1,T_2 \rangle$;}
		\ELSE
		\STATE{$T \gets \langle \text{Type}~\RN{3} \text{ leaf node} \rangle$;}
		\ENDIF
		\ENDIF
	\end{algorithmic}
	\label{Algo:bitmappartition}
\end{algorithm}

The {\em conquer} step accepts $s^*_{d^*}$ for partitioning if and only if bits are saved, i.e.,  $\mathcal{L}_{s^*_{d^*}}(\mathcal{M}) < \mathcal{L}(\mathcal{M})$. In that case, the the halves $\mathcal{M}_{s^*_{d^*},1}$ and $\mathcal{M}_{s^*_{d^*},2}$ are considered for further recursive splitting. Otherwise, it is classified as a Type~$\RN{3}$ leaf node (mixed).

The greedy BTBD heuristic is formally presented in Algorithm~\ref{Algo:bitmappartition}.

\subsection{Encoding and decoding}
\label{sec:Ch3_encoding}

For each frame, the encoder first compresses data maps $\mathcal{M}_{\text{div}_{64}}$, $\mathcal{M}_{\text{div}_{32}}$, $\mathcal{M}_{\text{div}_{16}}$, $\mathcal{M}_{\text{mode}}$,  $\mathcal{M}_{\overline {\text{MVZ}}}$, and $\mathcal{M}_{\text{res}}$ in order to preserve don't-care dependencies. Each data map is partitioned with BTBD and the associated partition-tree and leaf nodes are encoded with Huffman coding and CAAC, respectively. Finally, the encoder compresses the non-zero MV components.

\setlength\tabcolsep{8.5pt}
\begin{table}[!tb]
	\renewcommand{\arraystretch}{\tablerowstretch}
	\centering
	\caption{Huffman Codes of Partition-Tree Nodes}
	\label{tab:Ch3_huffmantreenode}
	\begin{tabular}{cccrr} \toprule
		\multirow{2}{*}{Node} & \multirow{2}{*}{Type} &    \multirow{2}{*}{Description} &    \multicolumn{2}{c}{Codeword} \\ \cmidrule(lr){4-5}
		& & & Bitmap & Intmap \\ \midrule[\heavyrulewidth]
		\multirow{3}{*}{Leaf} & $\RN{1}$ & all \num{0}'s & $\mathsf{00}$ & $\mathsf{001}$ \\
		& $\RN{2}$ & all same non-zero values & $\mathsf{1000}$ & $\mathsf{000}$ \\
		& $\RN{3}$ & mixed values & $\mathsf{101}$ & $\mathsf{01}$ \\ \midrule
		\multirow{3}{*}{Split} & X & along $x$-axis & $\mathsf{11}$ & $\mathsf{11}$ \\
		& Y & along $y$-axis & $\mathsf{01}$ & $\mathsf{10}$ \\
		& P & along $p$-axis & $\mathsf{1001}$ & N/A \\ \bottomrule
	\end{tabular}
\end{table}

\subsubsection{Partition-tree coding}
\label{sec:ch3_treeencoding}

The binary partition-tree is encoded following the pre-order depth-first traversal where the root of the tree is encoded first, then recursively the left sub-tree, and finally, recursively the right sub-tree. A tree can have six types of nodes, three types (X, Y, and P) of internal nodes representing binary split along the respective axis and three types ($\RN{1}, \RN{2}, \textrm{and\,} \RN{3}$) of leaf nodes representing coding blocks. These types are encoded using Huffman codes in  Table\,\ref{tab:Ch3_huffmantreenode} that are generated by analysing the typical probability of each type in multiview test sequences. For a bitmap, the split position of the internal nodes are also encoded using fixed-length codes (see Section~\ref{sec:BTBD}). For an intmap, however, the split positions are fixed and hence, no encoding is needed.

\subsubsection{Leaf node coding}
\label{sec:ch3_mixedleafcoding}

Once a partitioning-tree is encoded, the contents of its Type~$\RN{2}$ (intmap only) and $\RN{3}$ leaf nodes are encoded following the same tree-traversal order. For Type~$\RN{2}$ leaf nodes in an intmap, the SNZ value of each node is encoded using arithmetic coding over all such nodes.

Type~$\RN{3}$ leaf nodes are encoded with multiple coding modes to exploit varieties in probability distribution and the level of clustering tendency. Using partitioning or not, with context-adaptive or context-free arithmetic coding, differentiates four coding modes $\textsf{P}\textsf{C}$,  $\textsf{P}\bar{\textsf{C}}$, $\bar{\textsf{P}}\textsf{C}$, and  $\bar{\textsf{P}}\bar{\textsf{C}}$. In addition, using JPEG-LS without any partitioning is also available for $\mathcal{M}_{\text{res}}$ as the fifth coding mode $\textsf{J})$ to exploit 1D runs where no clustering is evident. For each map, the coding mode with the shortest code-length is selected and signalled using fixed-length \SI[number-unit-product = -]{2}{\bit} (all but residual maps) or \SI[number-unit-product = -]{3}{\bit} (residual maps) codes. For residual maps, the range $[0,R]$ is also signalled using an \SI[number-unit-product = -]{8}{\bit} unsigned integer.   

If a coding mode with CAAC is selected, different context model is used for nominal 2D bitmaps ($\mathcal{M}_{\text{div}_{64}}$, $\mathcal{M}_{\text{div}_{32}}$, and $\mathcal{M}_{\text{div}_{16}}$), the nominal 3D bitmap ($\mathcal{M}_{\overline {\text{MVZ}}}$), the nominal intmap ($\mathcal{M}_{\text{mode}}$), and the ordinal intmap ($\mathcal{M}_{\text{res}}$), as outlined in Section~\ref{sec:context}.

\subsubsection{Motion vector coding}

As zero-valued MV components in a frame have already been encoded with 2D bitmap $\mathcal{M}_{\overline {\text{MVZ}}}$, only the non-zero MV components need encoding in row-major scanning-order. Let $\omega$ be the search-width used in motion search. Then the non-zero MV components are drawn from a signed-integer source with $2\omega$ symbols $\in [-\omega,\omega] \setminus \{0\}$.

3D-HEVC encodes MV components independently using predictive coding to exploit that spatial correlations in neighbouring motion vectors. The median of already-encoded adjacent MV components is considered as the prediction and the difference $\epsilon_p \in [-2\omega,2\omega]$, $p \in \{1,2\}$, is encoded with Exp-Golomb codes of order-0 using $2\lceil \log_{2}(|\epsilon_p|+1)\rceil+1$\,\si{\bit}. This technique is used by the first coding mode $\textsf{PG}$. When $\epsilon_p$ follows TSG distribution, due to high spatial correlations, the Exp-Golomb code is optimal. To cover cases with weaker correlations, a second coding mode $\textsf{PA}$ is introduced, which uses arithmetic coding after signalling the range of symbols in $\lceil \log_2{2\omega} \rceil $\,\si{\bit}.

If predictive coding is avoided, the absence of zero-values itself can be exploited to achieve additional half a bit compression efficiency, on average, per non-zero MV component. This third coding mode $\bar{\textsf{P}}\textsf{G}$ uses modified Exp-Golomb codes of order-0 such that all positive or negative values, whichever is the majority in a frame, are encoded with codes of their preceding values of same sign (0 is assumed of both signs), after signalling the sign of majority in \SI{1}{\bit}. Similar to predictive coding, a fourth coding mode $\bar{\textsf{P}}\textsf{A}$ is introduced, which uses arithmetic coding after signalling the range of symbols in $\lceil \log_2{\omega} \rceil$\,\si{\bit}.

For each frame, the MV coding mode with the shortest code-length is selected and signalled using fixed-length \SI[number-unit-product = -]{2}{\bit} codes.

\subsubsection{Decoding}
\label{sec:ch3_decoding}

All operations used for encoding have inverse-operations. Hence, the decoding is straightforward, applying corresponding inverse-operations while maintaining the order used during encoding.

\section{Experimental Results and Analyses}
\label{sec:Ch3_result}

This section presents simulation results and analyses to demonstrate the superiority of the proposed hierarchical cuboid coding depth map sequence coder. After outlining the experimental setup (Section~\ref{sec:experimental_setup}), effectiveness of BTBD partitioning is demonstrated with some examples (Section~\ref{sec:BTBD_effictiveness}), and coding and view-synthesis results of BTBD and some contemporary coding techniques are presented (Section~\ref{sec:performance}).

\setlength\tabcolsep{5.5pt}
\begin{table}[!tb]
	\renewcommand{\arraystretch}{\tablerowstretch}
	\centering
	\caption{Multiview Video Plus Depth (MVD) Test Sequences}
	\label{tab:Ch1_tab1}
	\begin{tabular}{lccccc} \toprule
		\multirow{2}{*}{Sequence} & \multicolumn{3}{c}{Views} &
		Frame size \\ \cmidrule(lr){2-4} \cmidrule(lr){5-5}
		& Cameras & Selected & Synthesized & Frame rate \\ \midrule[\heavyrulewidth]
		
		$1$ \textit{Balloons}  & $7$ &   $1,3$ & $1.5,2,2.5$ & \\
		$2$ \textit{Newspaper} & $9$ & $2,4$ & $2.5,3,3.5$ & $1024 \times 768$p \\
		$3$ \textit{Lovebird1} & $12$ & $4,6$ & $4.5,5,5.5$ & $30$ fps \\
		$4$ \textit{Kendo} & $7$ & $1,3$ & $1.5,2,2.5$ & \\ \midrule
		$5$ \textit{PoznanStreet}  & $9$ & $5,3$ & $4.5,4,3.5$ & \\ 		
		$6$ \textit{PoznanHall2}  & $9$ &  $7,5$ & $6.5,6,5.5$ & $1920 \times 1088$p \\
		$7$ \textit{UndoDancer}  & CGI & $1,5$ & $2,3,4$ & $25$ fps \\
		$8$ \textit{GTFly}  & CGI &  $9,1$ & $7,5,3$ & \\ \bottomrule    
	\end{tabular}
\end{table}

\begin{figure*}[!tb]
	\centering
	
	\subfloat[]{\includegraphics[width=0.455\textwidth]{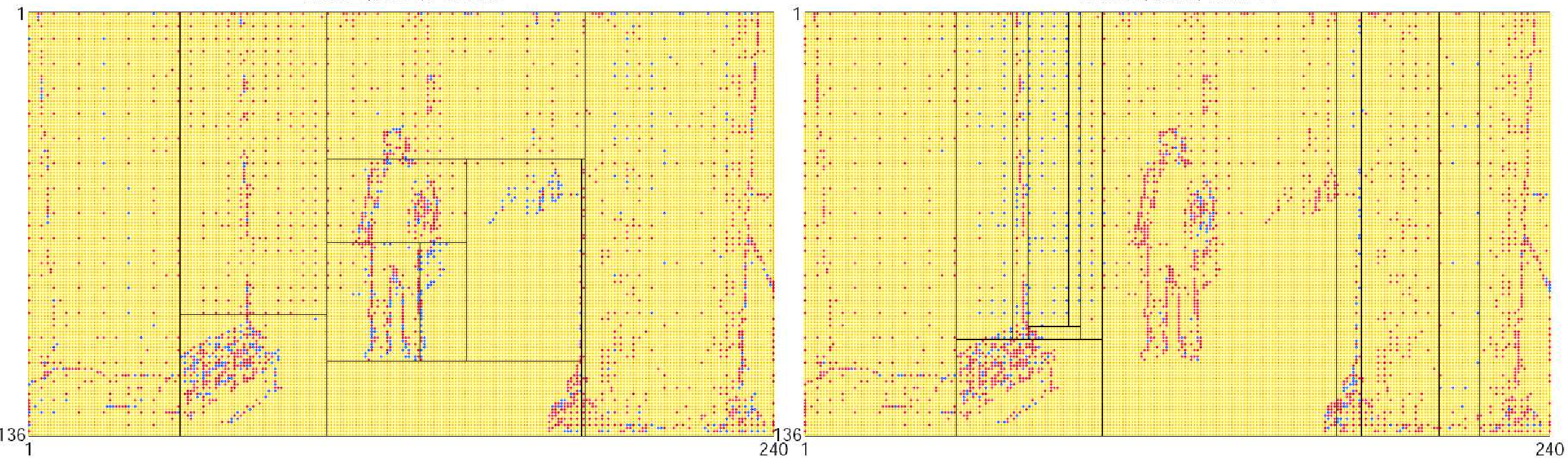}%
	\label{fig:ch3_figexample10a}}
	\hfill
	\subfloat[]{\includegraphics[width=0.27\textwidth]{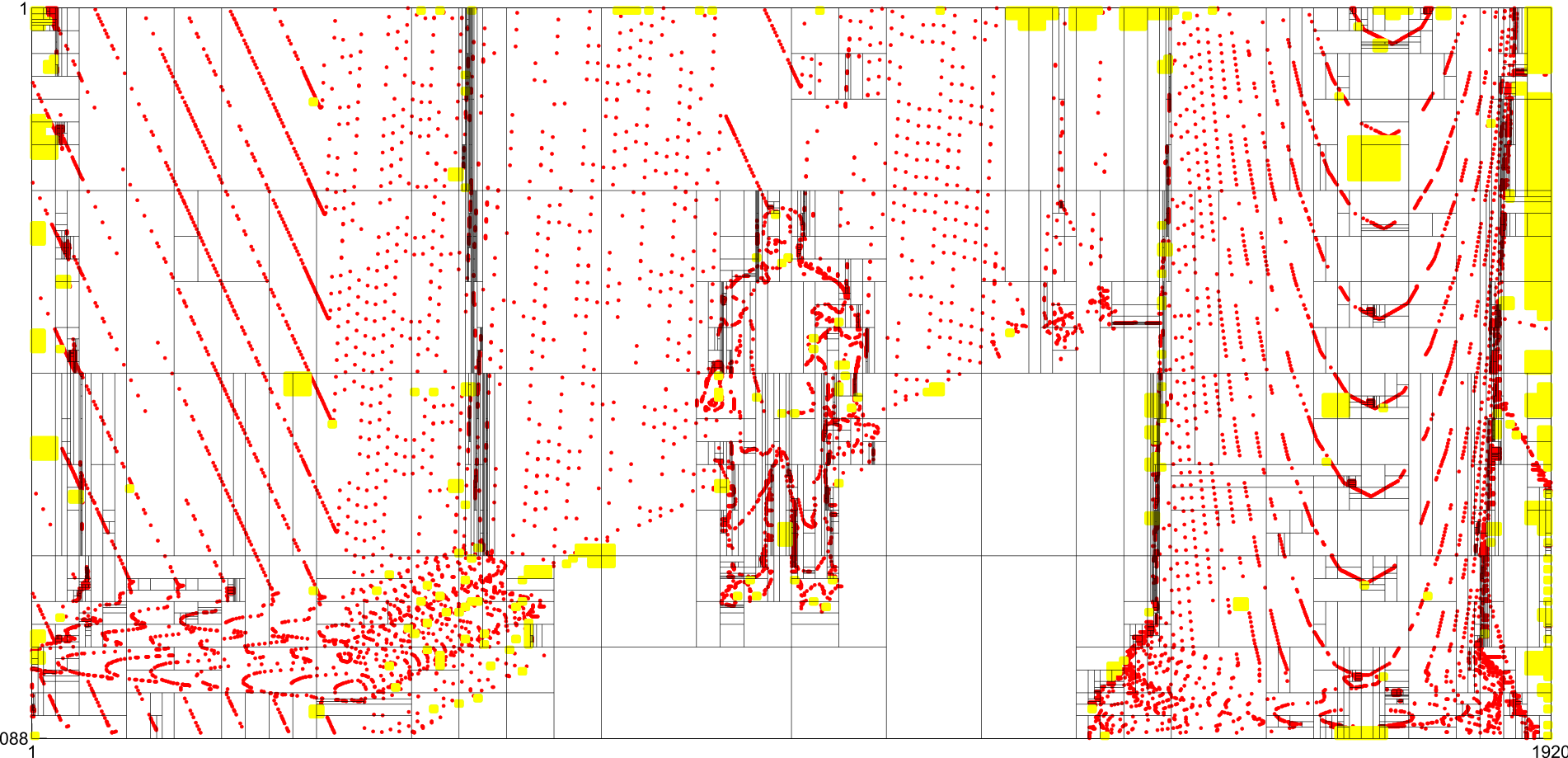}%
	\label{fig:ch3_figexample10b}}
	\hfill
	\subfloat[]{\includegraphics[width=0.255\textwidth]{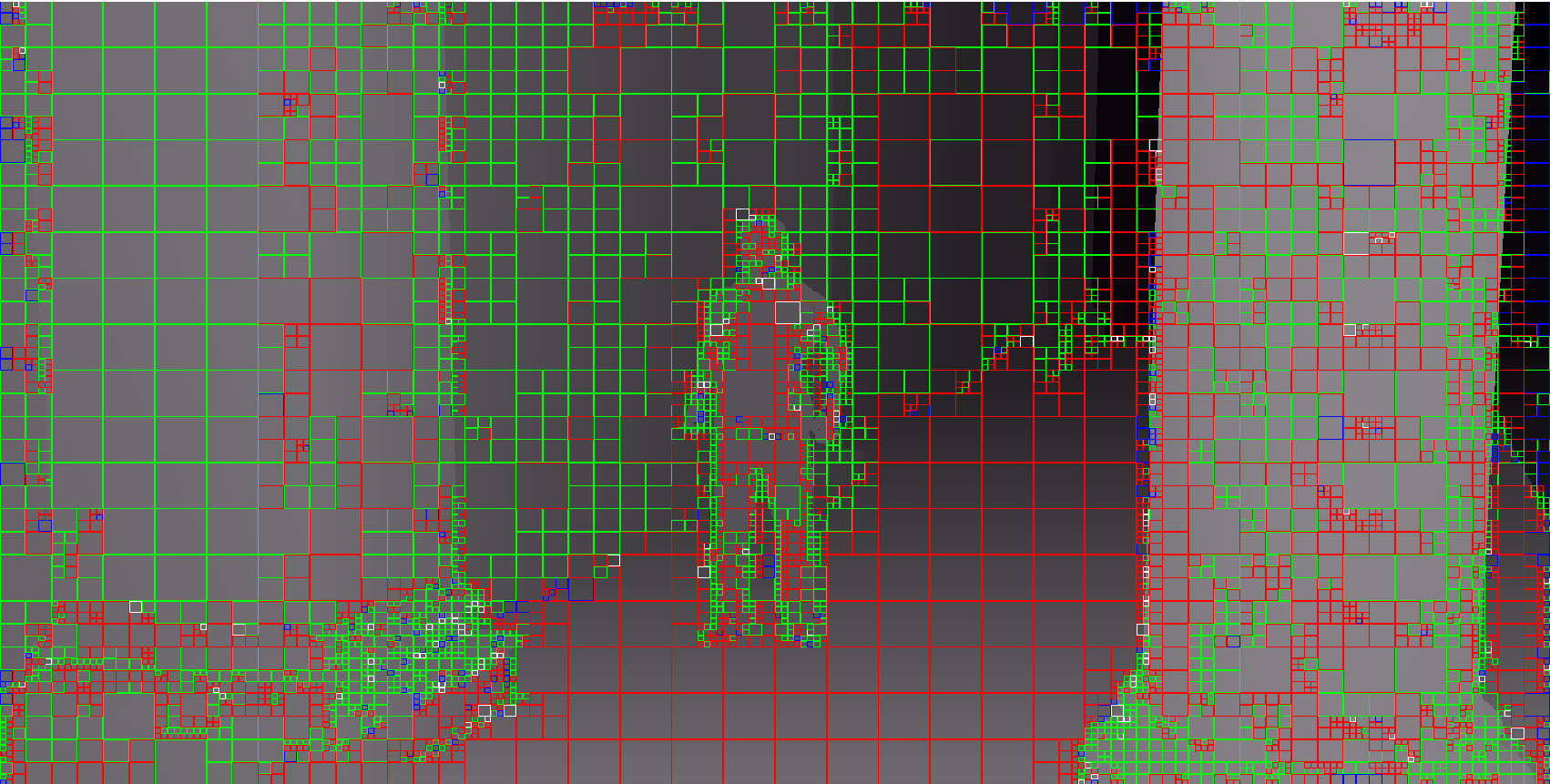}%
	\label{fig:ch3_figexample10c}}
	
	\caption{BTBD partitioning of (a) 3D bitmap $\mathcal{M}_{\overline {\text{MVZ}}}$, shown with two 2D bitmaps ($x$-components (left) and $y$-components (right)); (b) 2D residual intmap $\mathcal{M}_{\text{res}}$, where blue/white, red, and yellow represent zero, non-zero, and don't-care ($\dc$), respectively; and (c) CTU divisions with CU prediction modes, where red, blue, white and green represents \si{Intra}, \si{Skip}, \si{InterZ}, and \si{InterM} modes, respectively, of \textit{UndoDancer} sequence (view~1, frame~2, $Q = 1$). (best viewed in colour)}
	\label{fig:ch3_figexample10}
\end{figure*}

\subsection{Experimental Setup}
\label{sec:experimental_setup}

Experiments were carried out on eight MVD test sequences (Table\,\ref{tab:Ch1_tab1}) recommended by the 3D video coding standardization group~\cite{cfp3DVCT}. First six sequences were captured with linear camera arrangements of \numrange[range-phrase = --]{7}{12} cameras and the last two sequences were created with computer generated imagery (CGI). Texture frames were progressively scanned using $4$:$2$:$0$ colour sampling. For both texture and depth, the sample precision was restricted to \SI{8}{\bit}. Views of all sequences were rectified.

Performance of the proposed BTBD depth coder was compared against the state-of-the-art 3D-HEVC depth coder, using the 3D-HEVC Test Model (3D-HTM) reference software version 7.0~\cite{htm7}. Mono-view depth map sequences were encoded with BTBD and 3D-HEVC at different distortion levels by setting the scalar quantisation step $Q = 1, 3, \ldots, 15$ and the vector quantisation parameter $QP = 1, 5, 10, \ldots, 40$, respectively. The IPPPPPPPI group-of-picture (GOP) structure was used by both coders. With BTBD, P-frames were encoded through motion compensation using the diamond motion search~\cite{motionsearch2000} with search-width $\omega = 32$.

To evaluate near-lossless coding efficiency, depth maps of two selected views (Table\,\ref{tab:Ch1_tab1}) in each sequence were encoded independently. Reported coding results were obtained by averaging the bitrate (\si{bit~per~pixel} (\si{bpp})) and distortion (PSNR) of encoded depth maps of two coded-views.

3D-HEVC has not been designed for lossless compression as it is mostly based on explicit depth preservation with approximation. Thus, the following three alternatives were considered for conducting a comparative performance analysis of BTBD at lossless mode ($Q = 1$). Firstly, the pseudo-lossless 3D-HEVC ($QP = 1$), with the minimum possible quantisation noise, provides a baseline. Secondly, a simulated-lossless (Simul-LS) implementation of 3D-HEVC was considered. All edge-approximation coding modes were disabled and the residuals were encoded at frame-level similar to BTBD, i.e., without any frequency-domain transformation and using CAAC on ranked-residuals with BTBD context-model $\mathbb{C}_{\mathcal{M}_{\text{res}}}$ defined in (\ref{eq:ch3_contextres2}). Thirdly, to justify motion-compensation used in BTBD, each frame of a depth map sequence was encoded independently with JPEG-LS using the JPEG-LS reference software version 2.2~\cite{jpegsoft}. To evaluate lossless coding efficiency, the average bitrate (bpp) of the encoded depth maps of the first selected view (Table\,\ref{tab:Ch1_tab1}) and the corresponding compression-ratio $\textrm{cr} = 8/\textrm{bpp}$ are reported.

To evaluate view-synthesis quality, three intermediate equispaced virtual-views (Table\,\ref{tab:Ch1_tab1}) were synthesised between the two coded-views using their decoded depth maps. Reported  view-synthesis results were obtained by averaging the bitrate (bpp) of encoded depth maps of two coded-views and PSNR of rendered texture frames of three synthetic-views. The view synthesis reference software (VSRS) recommended by the 3D video coding standardization group~\cite{vsrs} was used. While analysing view-synthesis quality from depth maps that were encoded at various levels of distortion, any undue interference from the decoded lossy texture frames was avoided by keeping the texture quantisation level fixed. Throughout the simulations, decoded texture frames of 3D-HEVC at a fixed intermediate quantisation level $QP = 25$ of two coded-views were used with both BTBD and 3D-HEVC for all quantisation levels. In absence of any reference for calculating PSNR of the rendered texture frames at the virtual-viewpoints of synthetic views, virtual-references were synthesised using the original uncompressed texture frames and depth maps of the two encoded-views.

To compare the overall performance of BTBD and 3D-HEVC in the lossy-compression domain, along with their applications in view-synthesis, the average Bjøntegaard delta (BD) metrics \cite{BD} for bitrate (BD-BR) and distortion (BD-PSNR) were calculated. The logarithmic bitrate ($\log_{10} \textrm{bpp}$) vs distortion (PSNR) curves were interpolated with cubic polynomials and the difference between  their integrals over the common range was divided by the integration interval. Note that a negative BD-BR or a positive BD-PSNR means performance gain.

\subsection{Effectiveness of BTBD Partitioning}
\label{sec:BTBD_effictiveness}

The six data maps $\mathcal{M}_{\text{div}_{64}}$, $\mathcal{M}_{\text{div}_{32}}$, $\mathcal{M}_{\text{div}_{16}}$, $\mathcal{M}_{\text{mode}}$,  $\mathcal{M}_{\overline {\text{MVZ}}}$, and $\mathcal{M}_{\text{res}}$ that BTBD uses differ in terms of map-dimension (2D vs 3D), data-range (binary vs integer), and degrees of clustering tendency (high vs moderate). Due to space limitation, 3D bitmap $\mathcal{M}_{\overline {\text{MVZ}}}$ and 2D intmap $\mathcal{M}_{\text{res}}$ of \textit{UndoDancer} sequence (view~1, frame~2, $Q = 1$), covering all varieties, are selected to demonstrate the effectiveness of BTBD partitioning in exploiting the clustering tendency.

\subsubsection{Zero/non-zero MV component bitmap}

\figurename\,\ref{fig:ch3_figexample10a} presents BTBD partitioning of bitmap $\mathcal{M}_{\overline {\text{MVZ}}}$ of an arbitrary frame~2 in view~1 of \textit{UndoDancer} sequence for lossless compression. For viewing convenience, the 3D bitmap is shown with two 2D bitmaps of $x$- and $y$-components. The partitioning 
tree has one 1D, \num{23} 2D, and no 3D leaf nodes (cuboids) with maximum \num{663} and average \num{180} elements per cuboid. The number of zero, non-zero, and mixed leaf nodes are \numlist{1;1;22}, respectively. The number of splits orthogonal to $x$-, $y$-, and $p$-axis are \numlist{13;6;4}, respectively. The proposed BTBD technique encodes the bitmap with $\SI{2794}{\bit}$ and the non-zero MV components with  $\SI{11632}{\bit}$, using arithmetic coding ($\bar{\textsf{P}}\textsf{A}$ coding mode), in total $\SI{14426}{\bit}$. Without the bitmap, encoding all (zero and non-zero) MV components with the same coding mode would have required $\SI{22358}{\bit}$, resulting in $\SI{-35.5}{\percent}$ coding gain for BTBD.

\subsubsection{Residual intmap}
\label{sec:ch3_resmapresult}

Figs.~\ref{fig:ch3_figexample10b} and~\ref{fig:ch3_figexample10c} present BTBD partitioning of intmap $\mathcal{M}_{\text{res}}$ and the corresponding CTU divisions with CU prediction modes, respectively, of the same frame~2 in view~1 of \textit{UndoDancer} sequence for lossless compression. The partitioning tree of the residual intmap has \num{554} 1D and \num{1446} 2D leaf cuboids  with maximum \num{65216} and average \num{1017} elements per cuboid. The number of zero, all SNZ, and mixed leaf nodes are \numlist{833;49;1118}, respectively. CU prediction mode \si{InterM} is prominent as the sequence has high motion. For CU prediction mode \si{Skip}, the corresponding residuals are ignored, represented with the don't-care ($\dc$) symbols. Note that total \num{2000} leaf nodes is only \SI{6.1}{\percent} of the worst-possible CTU divisions when the frame is encoded with \num{32640} CUs of size \SI[product-units=single]{8 x 8}{pixel}. Therefore, the complexity overheads of the BTBD decoder is of little concern.

\setlength\tabcolsep{3.75pt}
\begin{table}[!tb]
	\renewcommand{\arraystretch}{\tablerowstretch}
	\centering
	\caption{Lossless Coding Performance of BTBD against JPEG-LS and 3D-HEVC (Pseudo- and Simulated-Lossless) where Compression-Ratio $\textrm{cr} = 8/\textrm{bpp}$} \label{tab:ch3_compression}
	\begin{tabular}{crrrrrrrr}
		\toprule
		\multirow{3}{*}{Seq} & \multicolumn{2}{c}{\multirow{2}{*}{JPEG-LS}} & \multicolumn{2}{c}{3D-HEVC} & \multicolumn{2}{c}{3D-HEVC} & \multicolumn{2}{c}{BTBD} \\
		& \multicolumn{2}{c}{} & \multicolumn{2}{c}{($QP = 1$)} & \multicolumn{2}{c}{(Simul-LS)} & \multicolumn{2}{c}{($Q = 1$)} \\ \cmidrule(lr){2-3} \cmidrule(lr){4-5} \cmidrule(lr){6-7} \cmidrule(lr){8-9}
		& \multicolumn{1}{c}{bpp} & \multicolumn{1}{c}{cr} & \multicolumn{1}{c}{bpp} & \multicolumn{1}{c}{cr} & \multicolumn{1}{c}{bpp} & \multicolumn{1}{c}{cr} & \multicolumn{1}{c}{bpp} & \multicolumn{1}{c}{cr} \\ \midrule[\heavyrulewidth]
		$1$ & $0.345$ & $\SI{23.2}[\times]{}$ & $0.852$ & $\SI{9.4}[\times]{}$ & $0.250$ & $\SI{32.0}[\times]{}$ & $\pmb{0.230}$ & $\pmb{\SI{34.8}[\times]{}}$ \\
		$2$ & $0.489$ & $\SI{16.4}[\times]{}$ & $0.714$ & $\SI{11.2}[\times]{}$ & $0.271$ & $\SI{29.5}[\times]{}$ & $\pmb{0.259}$ & $\pmb{\SI{30.9}[\times]{}}$ \\
		$3$ & $0.440$ & $\SI{18.2}[\times]{}$ & $0.192$ & $\SI{41.6}[\times]{}$ & $0.105$ & $\SI{76.3}[\times]{}$ & $\pmb{0.091}$ & $\pmb{\SI{87.7}[\times]{}}$ \\
		$4$ & $0.269$ & $\SI{29.8}[\times]{}$ & $0.837$ & $\SI{9.6}[\times]{}$ & $\pmb{0.268}$ & $\pmb{\SI{29.8}[\times]{}}$ & $\pmb{0.268}$ & $\pmb{\SI{29.8}[\times]{}}$ \\
		$5$ & $0.676$ & $\SI{11.8}[\times]{}$ & $0.567$ & $\SI{14.1}[\times]{}$ & $\pmb{0.291}$ & $\pmb{\SI{27.4}[\times]{}}$ & $0.298$ & $\SI{26.8}[\times]{}$ \\ 		
		$6$ & $0.137$ & $\SI{58.6}[\times]{}$ & $0.173$ & $\SI{46.4}[\times]{}$ & $0.144$ & $\SI{55.6}[\times]{}$ & $\pmb{0.097}$ & $\pmb{\SI{82.2}[\times]{}}$ \\
		$7$ & $0.324$ & $\SI{24.7}[\times]{}$ & $0.248$ & $\SI{32.3}[\times]{}$ & $0.089$ & $\SI{89.8}[\times]{}$ & $\pmb{0.074}$ & $\pmb{\SI{107.8}[\times]{}}$ \\
		$8$ & $0.270$ & $\SI{29.6}[\times]{}$ & $0.413$ & $\SI{19.4}[\times]{}$ & $0.231$ & $\SI{34.7}[\times]{}$ & $\pmb{0.199}$ & $\pmb{\SI{40.1}[\times]{}}$ \\ \midrule[\heavyrulewidth]
		Avg & $0.369$ & $\SI{21.7}[\times]{}$ & $0.499$ & $\SI{16.0}[\times]{}$ & $0.206$ & $\SI{38.8}[\times]{}$ & $\pmb{0.190}$ & $\pmb{\SI{42.2}[\times]{}}$ \\  \cmidrule(lr){2-3} \cmidrule(lr){4-5} \cmidrule(lr){6-7}
		Gain & \multicolumn{2}{c}{\SI{-48.6}{\percent}} & \multicolumn{2}{c}{\SI{-62.0}{\percent}} & \multicolumn{2}{c}{\SI{-8.0}{\percent}} \\ \bottomrule
	\end{tabular}
\end{table}

\subsection{Simulation Results}
\label{sec:performance}

\subsubsection{Lossless coding}

Table\,\ref{tab:ch3_compression} presents lossless coding performance results, depth bitrate (\si{bpp}) and corresponding compression-ratio, on all test sequences for JPEG-LS, 3D-HEVC (near- and simulated-lossless), and BTBD (lossless). Overall, average bitrate of \SIlist[list-units=single]{0.369;0.499;0.206;0.190}{bpp} was achieved by these four techniques, respectively. BTBD at lossless mode ($Q = 1$) outperformed other techniques with \SI{42.2}[\times]{} compression-ratio, on average, and \SIlist{-48.6;-62.0;-8.0}{\percent} coding gain against JPEG-LS, 3D-HEVC ($QP = 1$), and 3D-HEVC (Simul-LS), respectively. On individual sequences, BTBD ($Q = 1$) outperformed 3D-HEVC (Simul-LS) in all but two, \textit{Kendo} and \textit{PoznanStreet}, where clustering tendency are not sufficiently strong to justify any partitioning, especially the overhead of encoding the binary tree. For these two sequences, however, BTBD performed very close to 3D-HEVC (Simul-LS) with  negligible coding gain \SI{0.0}{\percent} and \SI{2.3}{\percent}, respectively.

Note that JPEG-LS outperformed 3D-HEVC ($QP = 1$) in five sequences despite being unable to exploit any temporal correlations. Unlike 3D-HEVC, JPEG-LS is capable of exploiting clustering tendency in some forms, e.g., compressing 1D-runs of very low values efficiently. This, however, is not surprising as performance gain by only intra-coding depth maps, based on geometric primitives, has been reported recently \cite{merkle2016}. Nevertheless, the inter-coding path with motion compensation has the upper hand in lossless/near-lossless coding once a mechanism of exploiting clustering tendency in multi-dimensions (1D, 2D, and 3D), e.g., the proposed BTBD partitioning, is introduced.

\setlength\tabcolsep{3.6pt}
\begin{table*}[!tb]
	\renewcommand{\arraystretch}{\tablerowstretch}
	\centering
	\caption{Near-Lossless Coding Performance of BTBD against 3D-HEVC with Average Bjøntegaard Delta Metrics for Depth Bitrate (BD-BR) and Depth Distortion (BD-PSNR)}
	\label{table:near-lossless}
	\begin{tabular}{cc|rrrrrrrrr|rrrrrrr|r}
		\toprule
		\multirow{2}{*}{Seq} & Rate & \multicolumn{9}{c|}{3D-HEVC ($QP = \,$)} & \multicolumn{7}{c|}{BTBD ($Q = \,$)} & \multicolumn{1}{c}{BD-BR} \\ \cmidrule(lr){2-2} \cmidrule(lr){3-11} \cmidrule(lr){12-18} \cmidrule(lr){19-19}
		& Dist & \multicolumn{1}{c}{$1$} & \multicolumn{1}{c}{$5$} & \multicolumn{1}{c}{$10$} & \multicolumn{1}{c}{$15$} & \multicolumn{1}{c}{$20$} & \multicolumn{1}{c}{$25$} & \multicolumn{1}{c}{$30$} & \multicolumn{1}{c}{$35$} & \multicolumn{1}{c|}{$40$} & \multicolumn{1}{c}{$3$} & \multicolumn{1}{c}{$5$} & \multicolumn{1}{c}{$7$} & \multicolumn{1}{c}{$9$} & \multicolumn{1}{c}{$11$} & \multicolumn{1}{c}{$13$} & \multicolumn{1}{c|}{$15$} & \multicolumn{1}{c}{BD-PSNR} \\ \midrule[\heavyrulewidth]
		\multirow{2}{*}{$1$} & \si{bpp} & $0.857$ & $0.638$ & $0.397$ & $0.207$ & $0.093$ & $0.038$ & $0.015$ & $0.007$ & $0.003$ & $0.167$ & $0.125$ & $0.101$ & $0.065$ & $0.040$ & $0.030$ & $0.026$ & \SI{-76.3}{\percent} \\
		& PSNR & $61.4$ & $58.1$ & $54.3$ & $51.0$ & $47.6$ & $44.6$ & $41.5$ & $39.2$ & $36.3$ & $59.9$ & $56.9$ & $53.8$ & $51.9$ & $50.3$ & $49.2$ & $48.8$ & \SI{6.42}{\dB} \\ 
		\multirow{2}{*}{$2$} & \si{bpp} & $0.764$ & $0.530$ & $0.313$ & $0.161$ & $0.074$ & $0.033$ & $0.015$ & $0.007$ & $0.004$ & $0.168$ & $0.108$ & $0.063$ & $0.041$ & $0.031$ & $0.025$ & $0.020$ & \SI{-84.2}{\percent} \\
		& PSNR & $57.4$ & $52.7$ & $49.3$ & $46.9$ & $44.4$ & $41.7$ & $39.7$ & $38.2$ & $36.1$ & $57.3$ & $53.3$ & $51.3$ & $49.9$ & $49.2$ & $48.8$ & $48.1$ & \SI{8.01}{\dB} \\ 
		\multirow{2}{*}{$3$} & \si{bpp} & $0.205$ & $0.142$ & $0.085$ & $0.049$ & $0.026$ & $0.012$ & $0.006$ & $0.004$ & $0.002$ & $0.042$ & $0.021$ & $0.016$ & $0.012$ & $0.010$ & $0.009$ & $0.008$ & \SI{-88.8}{\percent} \\
		& PSNR & $55.8$ & $54.6$ & $53.4$ & $52.0$ & $50.5$ & $48.7$ & $47.0$ & $45.3$ & $42.4$ & $58.5$ & $55.9$ & $54.8$ & $54.0$ & $53.5$ & $52.8$ & $52.4$ & \SI{5.72}{\dB} \\ 
		\multirow{2}{*}{$4$} & \si{bpp} & $0.754$ & $0.537$ & $0.324$ & $0.178$ & $0.090$ & $0.042$ & $0.018$ & $0.008$ & $0.003$ & $0.200$ & $0.129$ & $0.099$ & $0.058$ & $0.036$ & $0.028$ & $0.024$ & \SI{-69.8}{\percent} \\
		& PSNR & $60.8$ & $58.3$ & $55.1$ & $52.1$ & $49.1$ & $46.0$ & $42.8$ & $39.3$ & $34.8$ & $59.5$ & $57.1$ & $52.7$ & $52.4$ & $52.1$ & $50.8$ & $50.5$ & \SI{5.54}{\dB} \\
		\multirow{2}{*}{$5$} & \si{bpp} & $0.619$ & $0.399$ & $0.213$ & $0.051$ & $0.027$ & $0.014$ & $0.007$ & $0.003$ & $0.003$ & $0.119$ & $0.056$ & $0.033$ & $0.023$ & $0.017$ & $0.014$ & $0.011$ & \SI{-79.6}{\percent} \\
		& PSNR & $59.4$ & $56.5$ & $53.8$ & $49.1$ & $46.6$ & $44.7$ & $43.0$ & $40.8$ & $40.8$ & $56.6$ & $54.3$ & $53.4$ & $52.7$ & $52.2$ & $51.9$ & $51.5$ & \SI{5.98}{\dB} \\ 
		\multirow{2}{*}{$6$} & \si{bpp} & $0.180$ & $0.108$ & $0.056$ & $0.031$ & $0.016$ & $0.008$ & $0.004$ & $0.003$ & $0.001$ & $0.030$ & $0.011$ & $0.005$ & $0.003$ & $0.002$ & $0.001$ & $0.001$ & \SI{-52.0}{\percent} \\
		& PSNR & $63.8$ & $62.4$ & $60.3$ & $57.1$ & $54.3$ & $51.5$ & $48.9$ & $46.1$ & $43.0$ & $63.4$ & $58.8$ & $58.3$ & $57.9$ & $57.7$ & $57.5$ & $57.3$ & \SI{8.67}{\dB} \\
		\multirow{2}{*}{$7$} & \si{bpp} & $0.252$ & $0.146$ & $0.065$ & $0.031$ & $0.016$ & $0.009$ & $0.005$ & $0.003$ & $0.002$ & $0.016$ & $0.011$ & $0.008$ & $0.007$ & $0.006$ & $0.005$ & $0.005$ & \SI{-97.9}{\percent} \\
		& PSNR & $55.5$ & $54.9$ & $53.7$ & $51.0$ & $50.7$ & $49.7$ & $48.7$ & $46.4$ & $43.3$ & $62.9$ & $60.9$ & $59.3$ & $58.1$ & $57.4$ & $56.4$ & $55.8$ & \SI{10.04}{\dB} \\
		\multirow{2}{*}{$8$} & \si{bpp} & $0.411$ & $0.258$ & $0.134$ & $0.063$ & $0.027$ & $0.012$ & $0.007$ & $0.004$ & $0.002$ & $0.059$ & $0.035$ & $0.024$ & $0.019$ & $0.015$ & $0.013$ & $0.011$ & \SI{-86.8}{\percent} \\
		& PSNR & $58.2$ & $57.0$ & $55.2$ & $53.1$ & $50.5$ & $48.8$ & $47.7$ & $46.0$ & $41.6$ & $60.3$ & $57.5$ & $55.7$ & $55.0$ & $54.2$ & $54.1$ & $53.0$ & \SI{5.39}{\dB} \\ \midrule[\heavyrulewidth]
		\multirow{2}{*}{Avg} & \si{bpp} & $0.505$ & $0.345$ & $0.198$ & $0.096$ & $0.046$ & $0.021$ & $0.010$ & $0.005$ & $0.002$ & $0.100$ & $0.062$ & $0.044$ & $0.028$ & $0.020$ & $0.016$ & $0.013$ & \SI{-79.4}{\percent} \\
		& PSNR & $59.0$ & $56.8$ & $54.4$ & $51.5$ & $49.2$ & $47.0$ & $44.9$ & $42.7$ & $39.8$ & $59.8$ & $56.8$ & $54.9$ & $54.0$ & $53.3$ & $52.7$ & $52.2$ & \SI{6.98}{\dB} \\ \bottomrule 
	\end{tabular}
\end{table*}

\subsubsection{Near-lossless coding}

Table\,\ref{table:near-lossless} presents near-lossless coding performance results, depth bitrate (\si{bpp}), depth distortion (PSNR), and the average Bjøntegaard delta (BD) metrics BD-BR and BD-PSNR, for 3D-HEVC ($1 \le QP \le 40$) and BTBD ($3 \le Q \le 15$) on all test sequences. BTBD outperformed 3D-HEVC for all sequences with BD-BR \SIrange{-52.0}{-97.9}{\percent} and BD-PSNR \SIrange{5.39}{10.04}{\dB}, on average \SI{-79.4}{\percent} coding gain and \SI{6.98}{\dB} PSNR gain.

BTBD retained very high quality in decoded depth maps for all sequences. For $Q = 3, \ldots, 15$, on average, it achieved near-lossless PSNR \SIrange{59.8}{52.2}{\dB} against low bitrate \SIrange{0.100}{0.013}{bpp} with compression-ratio \SI{80.0}[\times]{} to \SI{608.8}[\times]{}. To retain comparable near-lossless PSNR \SIrange{59.0}{51.5}{\dB}, 3D-HEVC needed much higher bitrate \SIrange{0.505}{0.096}{bpp} with compression-ratio \SI{15.8}[\times]{} to \SI{83.1}[\times]{} for $QP = 1, \ldots, 15$.

\setlength\tabcolsep{2.9pt}
\begin{table*}[!tb]
	\renewcommand{\arraystretch}{\tablerowstretch}
	\centering
	\caption{Performance of BTBD against 3D-HEVC in View-Synthesis Applications with Average Bjøntegaard Delta Metrics for Depth Bitrate (BD-BR) and Synthetic-Texture Distortion (BD-PSNR)}
	\label{table:view-synthesis}
	\begin{tabular}{cc|rrrrrrrrr|rrrrrrrr|r}
		\toprule
		\multirow{2}{*}{Seq} & Rate & \multicolumn{9}{c|}{3D-HEVC ($QP = \,$)} & \multicolumn{8}{c|}{BTBD ($Q = \,$)} & \multicolumn{1}{c}{BD-BR} \\ \cmidrule(lr){2-2} \cmidrule(lr){3-11} \cmidrule(lr){12-19} \cmidrule(lr){20-20}
		& Dist & 
		\multicolumn{1}{c}{$1$} & \multicolumn{1}{c}{$5$} & \multicolumn{1}{c}{$10$} & \multicolumn{1}{c}{$15$} & \multicolumn{1}{c}{$20$} & \multicolumn{1}{c}{$25$} & \multicolumn{1}{c}{$30$} & \multicolumn{1}{c}{$35$} & \multicolumn{1}{c|}{$40$} & \multicolumn{1}{c}{$1$} & \multicolumn{1}{c}{$3$} & \multicolumn{1}{c}{$5$} & \multicolumn{1}{c}{$7$} & \multicolumn{1}{c}{$9$} & \multicolumn{1}{c}{$11$} & \multicolumn{1}{c}{$13$} & \multicolumn{1}{c|}{$15$} & \multicolumn{1}{c}{BD-PSNR} \\ \midrule[\heavyrulewidth]
		\multirow{2}{*}{$1$} & \si{bpp} & $0.857$ & $0.638$ & $0.397$ & $0.207$ & $0.093$ & $0.038$ & $0.015$ & $0.007$ & $0.003$ & $0.227$ & $0.167$ & $0.125$ & $0.101$ & $0.065$ & $0.040$ & $0.030$ & $0.026$ & \SI{-30.0}{\percent} \\
		& PSNR & $45.2$ & $45.1$ & $45.0$ & $44.7$ & $44.1$ & $43.5$ & $43.0$ & $42.4$ & $41.8$ & $45.2$ & $45.0$ & $44.7$ & $44.2$ & $43.9$ & $43.7$ & $43.6$ & $43.4$ & \SI{0.14}{\dB} \\
		\multirow{2}{*}{$2$} & \si{bpp} & $0.764$ & $0.530$ & $0.313$ & $0.161$ & $0.074$ & $0.033$ & $0.015$ & $0.007$ & $0.004$ & $0.270$ & $0.168$ & $0.108$ & $0.063$ & $0.041$ & $0.031$ & $0.025$ & $0.020$ & \SI{-94.2}{\percent} \\
		& PSNR & $41.5$ & $40.7$ & $40.1$ & $39.5$ & $39.2$ & $38.3$ & $37.8$ & $37.7$ & $37.5$ & $43.3$ & $42.4$ & $41.7$ & $41.4$ & $41.3$ & $41.1$ & $40.9$ & $40.8$ & \SI{2.75}{\dB} \\
		\multirow{2}{*}{$3$} & \si{bpp} & $0.205$ & $0.142$ & $0.085$ & $0.049$ & $0.026$ & $0.012$ & $0.006$ & $0.004$ & $0.002$ & $0.093$ & $0.042$ & $0.021$ & $0.016$ & $0.012$ & $0.010$ & $0.009$ & $0.008$ & \SI{-8.2}{\percent} \\
		& PSNR & $42.9$ & $42.8$ & $42.5$ & $42.2$ & $41.8$ & $41.5$ & $41.3$ & $41.1$ & $40.6$ & $43.0$ & $42.2$ & $41.8$ & $41.5$ & $41.4$ & $41.3$ & $41.2$ & $41.1$ & \SI{0.03}{\dB} \\
		\multirow{2}{*}{$4$} & \si{bpp} & $0.754$ & $0.537$ & $0.324$ & $0.178$ & $0.090$ & $0.042$ & $0.018$ & $0.008$ & $0.003$ & $0.233$ & $0.200$ & $0.129$ & $0.099$ & $0.058$ & $0.036$ & $0.028$ & $0.024$ & \SI{-49.4}{\percent} \\
		& PSNR & $45.9$ & $45.8$ & $45.8$ & $45.6$ & $45.4$ & $45.2$ & $45.0$ & $44.6$ & $44.0$ & $45.9$ & $45.8$ & $45.7$ & $45.6$ & $45.5$ & $45.5$ & $45.4$ & $45.4$ & \SI{0.17}{\dB} \\
		\multirow{2}{*}{$5$} & \si{bpp} & $0.619$ & $0.399$ & $0.213$ & $0.051$ & $0.027$ & $0.014$ & $0.007$ & $0.003$ & $0.003$ & $0.313$ & $0.119$ & $0.056$ & $0.033$ & $0.023$ & $0.017$ & $0.014$ & $0.011$ & \SI{-29.8}{\percent} \\
		& PSNR & $41.9$ & $41.6$ & $41.1$ & $40.2$ & $39.8$ & $39.5$ & $39.3$ & $38.8$ & $38.8$ & $42.1$ & $40.8$ & $40.3$ & $40.0$ & $39.9$ & $39.8$ & $39.8$ & $39.7$ & \SI{0.20}{\dB} \\
		\multirow{2}{*}{$6$} & \si{bpp} & $0.180$ & $0.108$ & $0.056$ & $0.031$ & $0.016$ & $0.008$ & $0.004$ & $0.003$ & $0.001$ & $0.126$ & $0.056$ & $0.026$ & $0.013$ & $0.007$ & $0.004$ & $0.002$ & $0.002$ & \SI{-26.5}{\percent} \\
		& PSNR & $44.1$ & $43.9$ & $43.8$ & $43.5$ & $43.3$ & $43.0$ & $42.7$ & $42.4$ & $42.1$ & $44.2$ & $43.7$ & $43.4$ & $43.3$ & $43.1$ & $43.0$ & $42.9$ & $42.7$ & \SI{0.18}{\dB} \\
		\multirow{2}{*}{$7$} & \si{bpp} & $0.252$ & $0.146$ & $0.065$ & $0.031$ & $0.016$ & $0.009$ & $0.005$ & $0.003$ & $0.002$ & $0.074$ & $0.016$ & $0.011$ & $0.008$ & $0.007$ & $0.006$ & $0.005$ & $0.005$ & \SI{18.6}{\percent} \\
		& PSNR & $40.2$ & $40.0$ & $39.9$ & $39.8$ & $39.7$ & $39.6$ & $39.4$ & $39.0$ & $38.5$ & $40.5$ & $39.8$ & $39.3$ & $39.0$ & $38.8$ & $38.7$ & $38.4$ & $38.4$ & \SI{0.17}{\dB} \\
		\multirow{2}{*}{$8$} & \si{bpp} & $0.411$ & $0.258$ & $0.134$ & $0.063$ & $0.027$ & $0.012$ & $0.007$ & $0.004$ & $0.002$ & $0.197$ & $0.059$ & $0.035$ & $0.024$ & $0.019$ & $0.015$ & $0.013$ & $0.011$ & \SI{68.7}{\percent} \\
		& PSNR & $42.4$ & $42.3$ & $42.1$ & $41.9$ & $41.7$ & $41.5$ & $41.5$ & $41.4$ & $41.1$ & $42.5$ & $42.0$ & $41.5$ & $41.3$ & $41.0$ & $40.7$ & $40.4$ & $40.1$ & \SI{-0.20}{\dB} \\ \midrule[\heavyrulewidth]
		\multirow{2}{*}{Avg} & \si{bpp} & $0.505$ & $0.345$ & $0.198$ & $0.096$ & $0.046$ & $0.021$ & $0.010$ & $0.005$ & $0.002$ & $0.191$ & $0.103$ & $0.064$ & $0.045$ & $0.029$ & $0.020$ & $0.016$ & $0.013$ & \SI{-18.9}{\percent} \\
		& PSNR & $43.0$ & $42.8$ & $42.5$ & $42.2$ & $41.9$ & $41.5$ & $41.2$ & $40.9$ & $40.6$ & $43.3$ & $42.7$ & $42.3$ & $42.0$ & $41.9$ & $41.7$ & $41.6$ & $41.5$ & \SI{0.43}{\dB}
		\\ \bottomrule 
	\end{tabular}
\end{table*}

\subsubsection{View-synthesis}

Table\,\ref{table:view-synthesis} presents performance results in view-synthesis applications, depth bitrate (\si{bpp}), distortion in synthetic-texture (PSNR), and the average BD-BR and BD-PSNR metrics, for 3D-HEVC ($1 \le QP \le 40$) and BTBD ($1 \le Q \le 15$) on all test sequences. Compared to 3D-HEVC, BTBD rendered better quality synthetic-views for the first six sequences with BD-BR \SIrange{-8.2}{-94.2}{\percent} and BD-PSNR \SIrange{0.03}{2.75}{\dB}. For the remaining two CGI sequences \textit{UndoDancer} and \textit{GTFly}, BTBD performed mixed (BD-BR unfavourable; but BD-PSNR favourable) and unfavourably, respectively. On average, BTBD achieved  \SI{-18.9}{\percent} coding gain and \SI{0.43}{\dB} PSNR gain against 3D-HEVC.

\begin{figure}[!tb]
	\centering
	
	\subfloat{\includegraphics[width=0.49\columnwidth]{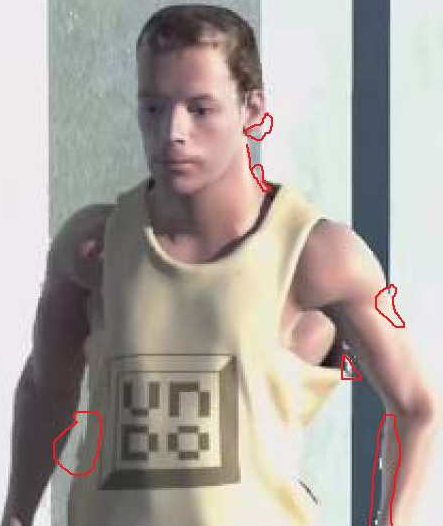}}
	\hfill
	\subfloat{\includegraphics[width=0.49\columnwidth]{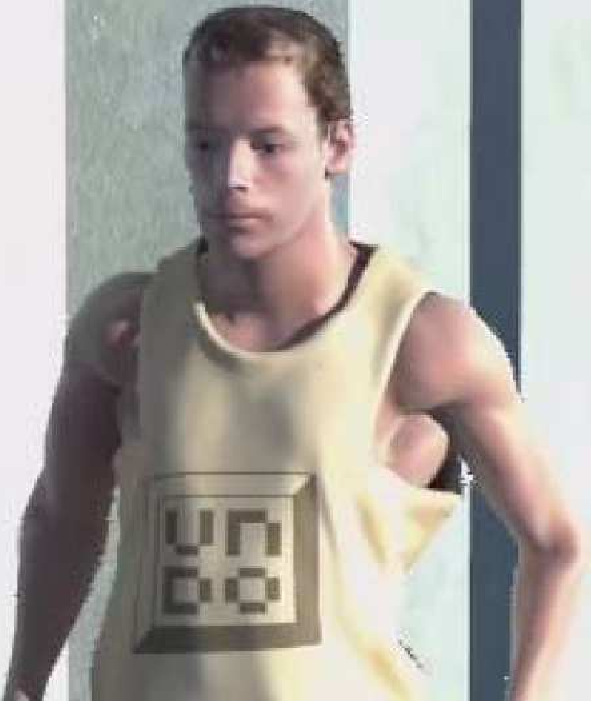}}
	
	\caption{Synthetic-view~4 (frame~5) of \textit{UndoDancer} sequence from the decoded depth maps of (left) 3D-HEVC ($QP = 40$), where shape deformations are encircled in red, and (right) BTBD ($Q = 15$), both rendered at the same synthetic-texture PSNR \SI{39}[\simeq]{\dB}. (best viewed in colour)}
	\label{fig:Ch3_exampledancerdep1}
\end{figure}

For natural sequences, depth maps are estimated indirectly from rectified texture frames at two or more views using stereo matching techniques. The larger the matching-window size, the smoother the depth map with less details. Depth maps of CGI sequences, however, are obtained directly from the 3D model of the scene geometry. They retain finer details with less smoothness, leading to exhibiting weaker clustering tendency. This explains why BTBD partitioning was less effective on CGI sequences. Nevertheless, subjective evaluations reveal that compared to 3D-HEVC, BTBD rendered views are visually more appealing due to lower shape deformation. \figurename\,\ref{fig:Ch3_exampledancerdep1} provides an example for subjective evaluation of synthetic-views of \textit{Undodancer} sequence that were generated from the decoded depth maps of same quality by both techniques. Close scrutiny of the synthetic-view from 3D-HEVC depth maps reveals a number of shape deformations (encircled in red) due to use of edge-approximation modes. The synthetic-view from BTBD depth maps is free from such flaws.

\section{Conclusion}
\label{sec:Ch3_conclusion}

In this paper, a novel independent depth map sequence coder has been developed, which can efficiently exploit high spatio-temporal and inter-component correlations in CTU divisions, CU prediction modes, motion vectors, and residuals with 2D/3D binary/integer data maps. A simple but highly effective greedy heuristic based hierarchical decomposition technique has been introduced to adaptively partition the data maps into cuboids of data with skewed probability distribution that are encoded independently with context-adaptive arithmetic coding. The proposed BTBD technique achieved \SI{42.2}[\times]{} compression-ratio on average for lossless coding to outperform the pseudo- ($QP = 1$) and simulated-lossless 3D-HEVC with average coding gain \SIlist{-62.0;-8.0}{\percent}, respectively. For near-lossless coding, BTBD performed far superior compared to 3D-HEVC with average \SI{-79.4}{\percent} coding gain and \SI{6.98}{\dB} PSNR gain. By restricting quantisation to spatial-domain, edges are preserved inherently. The synthetic-views rendered from the decoded depth maps from BTBD had superior visual quality with less shape deformations, achieving on average \SI{0.43}{\dB} PSNR gain against 3D-HEVC, which is perceptually significant. The proposed BTBD coding technique is elegant and it may have applications in other domains such as texture coding, hyperspectral image coding, image segmentation, and compressed-domain image processing that will be investigated in future.

\IEEEtriggeratref{42}
\bibliographystyle{IEEEtran}
\bibliography{IEEEabrv,bare_jrnlV1}

\end{document}